# Thermodynamic–Kinetic Decoupling Enables Stable Excitonic Emission in Defect-Tolerant Cu-Based Quantum Dots

*Haoran Chen,*[a] *Zhipeng Xu,*[a] *Chunjian Li,*[a] *Lei Hou,*[b,*] *Dechao Yu,*[b] *Xiaobin Xie,*[c] *Yue Liu,*[d] *Bohua Dong,*[a] *Lixin Cao*[a,*] *and Chenghui Xia*[a,*]

a. Engineering Research Center of Marine Materials and Protection Technology, Ministry of Education, Key Laboratory of Marine Equipment Materials and Protection of Shandong Province, School of Materials Science and Engineering, Ocean University of China, No. 1299, Sansha Road, Huangdao District, Qingdao, 266000, China
b. Engineering Research Center of Optical Instrument and System, Ministry of Education and Shanghai Key Lab of Modern Optical System, University of Shanghai for Science and Technology, Shanghai 200093, China
c. Guangdong Provincial Key Laboratory of Optical Information Materials and Technology & Institute of Electronic Paper Displays, South China Academy of Advanced Optoelectronics, South China Normal University, Guangzhou 510006, China
d. College of Chemistry, Sichuan University, Chengdu 610064, China

*Corresponding authors:

E-mails: h_lei@pku.edu.cn; caolixin@ouc.edu.cn; c.xia@ouc.edu.cn

## Abstract

Colloidal quantum dots that simultaneously offer room-temperature single-photon purity and high photoluminescence quantum yield are sought for quantum optics, but remain elusive in environmentally benign materials. We introduce a thermodynamic–kinetic decoupling strategy that transforms defect-tolerant $CuInS_2$ quantum dots into bright, narrowband, and photostable single-photon emitters. $Zn^{2+}$ alloying strains the lattice, thermodynamically suppressing native copper vacancies and narrowing the emission from a broad defect band (~300 meV) to an excitonic line (~120 meV). $Ga^{3+}$ incorporation then kinetically pins the cation sublattice against $Cu^{+}$ migration, preventing defect regeneration during ZnS shell growth. The resulting Cd-free core/shell dots achieve near-unity quantum yield (~98%) while retaining narrow excitonic emission. Critically, room-temperature single-dot spectroscopy reveals homogeneous linewidths as low as ~58 meV, strongly suppressed blinking, and high-purity single-photon emission ($g^2(0) = 0.06$). This stabilized excitonic emission directly reduces reabsorption losses in luminescent solar concentrators, yielding an external optical efficiency of 12.68%. Our work establishes a generalizable framework to unlock intrinsic excitonic photophysics in ion-mobile, defect-prone semiconductors, opening a viable path toward high-performance heavy-metal-free emitters for quantum light sources.

Colloidal quantum dots (QDs) based on heavy-metal-free semiconductors are widely regarded as key building blocks for sustainable optoelectronic technologies[1-6]. Among them, copper indium sulfide (CIS) QDs are particularly appealing because their emission can be tuned across the deep-red to near-infrared region while retaining compatibility with low-cost, solution-processed device architectures[7-9]. Over the past decade, significant advances have been achieved in improving their photoluminescence quantum yields (PLQYs), which in many cases now approach unity[10-12]. In contrast, the spectral characteristics of CIS QDs have proven far more resistant to improvement. Their emission remains intrinsically broad, typically exceeding ~300 meV, indicating that radiative recombination is still dominated by defect-related states rather than by band-edge excitons[13,14]. This persistent lack of stable, narrow-band emission has limited the deployment of CIS QDs in applications that require high color purity, optical coherence, and reproducible single-particle light-emitting behavior.

The difficulty of spectral narrowing in CIS QDs is closely tied to the defect-tolerant nature of multinary I–III–VI semiconductors. Copper vacancies and related antisite defects form readily and introduce a high density of electronic states within the bandgap[15-17]. As a result, excited carriers are efficiently captured by these states, leading to broad and strongly Stokes-shifted photoluminescence (PL). A wide range of material design strategies, including shell passivation[18-20], cation alloying[21,22], and core/shell heterostructure engineering[23,24], have been developed to mitigate nonradiative recombination and to enhance PLQY. While these approaches have been highly effective in suppressing nonradiative losses, their impact on the intrinsic linewidth has been modest, and reports of genuine band-edge emission remain scarce and often lack long-term stability. Taken together, these results suggest that controlling defect populations solely through their formation energetics is insufficient to achieve durable spectral purity.

An additional and less systematically addressed challenge arises from the dynamic nature of defects in CIS QDs. Copper ions exhibit unusually high mobility within the crystal lattice[25-28], particularly under the thermal and chemical conditions encountered during post-synthetic processing, such as shell growth or annealing. Even when defect densities are initially reduced through alloying or passivation, this mobility enables defects to re-emerge, leading to renewed spectral broadening and degradation of optical coherence[29-31]. From this perspective, the long-standing difficulty in realizing narrow-band emission in CIS QDs reflects the combined influence of two factors: the thermodynamic tendency toward defect formation and the kinetic propensity for defect regeneration. These two aspects are often entangled in practice and have rarely been addressed in a coordinated manner.

In this work, we demonstrate that intrinsic and stable excitonic emission in CIS QDs can be achieved by concurrently addressing defect formation and defect regeneration through a unified thermodynamic–kinetic design strategy. Controlled $Zn^{2+}$ alloying introduces lattice strain that raises the formation energy of deep defects and shifts recombination toward the band-edge, while subsequent incorporation of $Ga^{3+}$ kinetically suppresses $Cu^{+}$ migration and stabilizes the optimized defect landscape during shell growth. As a result, the resulting cadmium-free CIS-based QDs exhibit narrow excitonic emission, near-unity PLQY after epitaxial ZnS passivation, and excellent spectral and photostability. Single-particle measurements further confirm suppressed blinking and homogeneous linewidths. More broadly, this work points to a general guideline for improving the light-emitting performance of defect-prone multinary semiconductors: durable access to intrinsic optical properties requires the simultaneous engineering of defect energetics and kinetics. Importantly, this strategy does not rely on specific chemical identities, but on identifying the dominant defect species and the most mobile ionic sublattice, and independently engineering their thermodynamic stability and kinetic accessibility.

## Results

### Thermodynamic suppression of defect formation via $Zn^{2+}$ alloying

The PL of wurtzite CIS QDs is characteristically broad and strongly Stokes-shifted, reflecting a recombination landscape dominated by intrinsic defect states that form readily in defect-tolerant I–III–VI semiconductors[32,33]. A central challenge is therefore to thermodynamically suppress the formation of the dominant native defects, most notably Cu vacancies ($V_{Cu}$), so that radiative recombination can be redirected toward a narrower, exciton-like channel. We implement this thermodynamic lever by incorporating $Zn^{2+}$ into CIS via a controlled cation-exchange alloying process, producing homogeneous Cu–In–Zn–S (CIZS) QDs (Fig. 1a).

$Zn^{2+}$ alloying immediately reshapes the ensemble optical response. Compared with CIS, CIZS exhibits a pronounced narrowing of the emission and an exciton-like spectral profile (Fig. 1b), indicating substantial suppression of the broad defect-dominated contribution. Importantly, the alloying chemistry provides a handle to tune the competition between the narrowed band (Band-*N*) and the residual broad defect band (Band-*B*). The photoluminescence excitation spectra monitored at Band-*N* and Band-*B* exhibit nearly identical profiles, indicating they share the same excited-state manifold; the higher PLE intensity of Band-*N* suggests a larger emission branching ratio compared to Band-*B* (Supplementary Fig. 3). As shown in Fig. 1c, varying the Zn halide precursor ($ZnI_2$, $ZnBr_2$, $ZnCl_2$) systematically modulates the relative prominence of these two components, establishing that the narrowed emission is a controlled outcome of Zn incorporation rather than a coincidental spectral fluctuation.

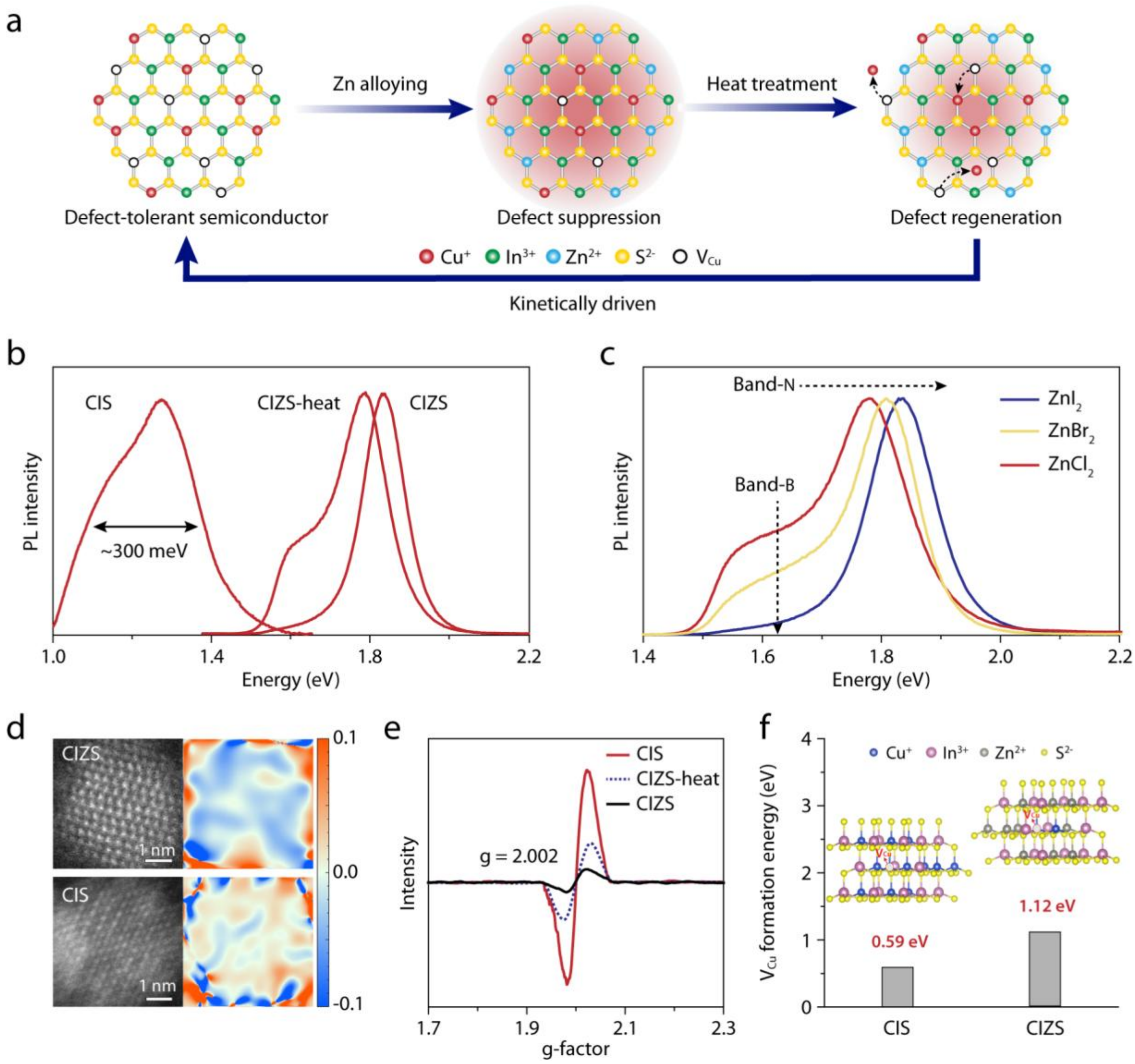

**Figure 1. $Zn^{2+}$ alloying thermodynamically suppresses native defects and redirects CIS emission toward an exciton-like narrowed band. (a)** Schematic of thermodynamic defect suppression and kinetic defect locking (strain-and-pin design). **(b)** PL spectra of CIS, CIZS and the CIZS QDs after heating at 230 °C for 2 h. **(c)** PL spectra of CIS QDs after $Zn^{2+}$ alloying using different Zn precursors at 200 °C for 1 h. **(d)** High-resolution HAADF-STEM images and corresponding geometric phase analysis (GPA) of CIS and CIZS QDs. **(e)** EPR spectra of CIS, CIZS and the CIZS QDs after heating at 230 °C for 2 h. **(f)** Cu vacancy formation energy in CIS versus CIZS QDs.

By monitoring the $ZnI_2$ alloying process in real time (Supplementary Fig. 4), we observed a systematic blue shift in the PL spectrum, which ultimately transitioned from broad defect emission to a symmetric, narrowed emission band. The full width at half maximum (FWHM) decreased sharply to 120 meV, while the PLQY increased to 57% (see Supplementary Method 1 for calculation details). Time-resolved photoluminescence measurements revealed a reduction in the average lifetime from 192 ns to 129 ns (Supplementary Fig. 5), indicating that fast band-edge radiative recombination has replaced slow defect-state recombination as the dominant emission pathway. Supplementary Fig. 6 reveals an optimal Zn-alloying window: increasing Zn/Cu up to ~3 suppresses defect emission and narrows the PL (minimum FWHM ~123 meV), whereas further Zn addition re-broadens the band, consistent with disorder/inhomogeneous broadening beyond the defect-suppression regime.

Structural evidence supports lattice contraction as the physical basis for this thermodynamic perturbation. High-resolution TEM and strain mapping reveal that Zn incorporation introduces a measurable compressive strain field relative to the CIS seeds (Fig. 1d and Supplementary Fig. 7). To further confirm the structural integrity, fast Fourier transform analysis was performed on selected regions. The resulting diffraction patterns can be consistently indexed to the characteristic features of the hexagonal wurtzite structure along the [100] and [101] zone axes, confirming phase uniformity and crystallographic consistency of the alloyed QDs (Supplementary Fig. 8). This strain is not presented as a structural curiosity. Rather, it provides a plausible physical mechanism by which the energetics of native defect formation can be altered in a soft, defect-tolerant lattice.

The XRD diffraction peaks of CIZS exhibit a systematic shift toward larger angles compared to CIS, indicating a reduction in interplanar spacing after alloying (Supplementary Fig. 9 and Supplementary Table 2). Meanwhile, TEM statistical analysis reveals that the average size of CIZS QDs is slightly smaller than that of CIS QDs (5.0 nm versus 5.1 nm), suggesting that the alloying process may involve subtle lattice restructuring or surface reconstruction, further supporting the regulatory role of lattice strain on defect distribution (Supplementary Fig. 10). Compositional analysis reveals that during $Zn^{2+}$ alloying, Zn ions predominantly occupy Cu sites. This selective occupation arises from the similarities in ionic radius and coordination chemistry between $Zn^{2+}$ and $Cu^{+}$, which facilitates their incorporation into the copper sublattice rather than indium sites. Consequently, while the overall lattice structure is maintained, local compressive strain is introduced (Supplementary Table 3). Consistent with this picture, electron paramagnetic resonance (EPR) measurements show a substantial suppression of the defect-associated signal ($g \approx 2.002$) in CIZS relative to CIS, while thermal treatment partially restores this signature (CIZS-heat, Fig. 1e), directly linking the narrowed emission to a reduced defect population under thermodynamic control.

First-principles calculations provide quantitative support for this thermodynamic interpretation (see Supplementary Method 2 for calculation details). The calculated formation energy of $V_{Cu}$ increases from 0.59 eV in CIS to 1.12 eV in CIZS (Fig. 1f), indicating that Zn-induced lattice perturbation raises the thermodynamic cost of the dominant native acceptor defect and thereby lowers its equilibrium concentration[34,35]. Together, the optical evolution, precursor-dependent control of the Band-*N*/Band-*B* balance, strain signatures, and defect-sensitive EPR response converge on a consistent conclusion. $Zn^{2+}$ alloying establishes a thermodynamically improved, low-defect baseline that enables spectral narrowing toward an exciton-like channel. However, the partial recovery of defect signatures upon heating (Fig. 1b,e) also foreshadows an additional limitation beyond formation energetics, motivating a subsequent strategy to kinetically stabilize the optimized cation configuration against defect regeneration.

**Kinetic stabilization of the defect landscape by $Ga^{3+}$ incorporation**

While $Zn^{2+}$ alloying thermodynamically suppresses the formation of native defects, it does not eliminate a second, intrinsically kinetic limitation of CIS-based QDs, i.e., the high mobility of $Cu^{+}$ ions in the soft, defect-tolerant lattice[36]. Under thermal or chemical perturbations, most notably during post-synthetic shell growth, this mobility enables cation rearrangement and defect regeneration, which can reintroduce deep trap states even when their equilibrium concentration is reduced. Durable spectral narrowing therefore requires not only a thermodynamically improved defect landscape, but also kinetic stabilization against $Cu^{+}$ migration.

To address this limitation, we introduce $Ga^{3+}$ into the Zn-alloyed CIS lattice, forming Cu–In–Zn–Ga–S (CIZGS) QDs. X-ray photoelectron spectroscopy confirms the successful incorporation of Ga in the +3 oxidation state, with no evidence of segregated gallium oxide phases (Fig. 2a, see Supplementary Fig. 11 for other elements). Importantly, $Ga^{3+}$ incorporation preserves the narrow emission linewidth established by $Zn^{2+}$ alloying (~123–124 meV) while substantially increasing the PLQY from ~57% to ~84% prior to shell growth (Fig. 2b). This selective improvement in efficiency, without further linewidth reduction, indicates that $Ga^{3+}$ does not primarily act as an additional thermodynamic suppressor of defect formation, but instead enhances the kinetic stability of the recombination landscape. Supplementary Fig. 12 further supports this interpretation: excessive Ga incorporation leads to a noticeable drop in PLQY. This trend indicates that Ga primarily acts as a kinetic 'locking' agent, stabilizing the optimized cation configuration—delivering high PLQY without additional spectral narrowing—whereas over-doping with Ga introduces new nonradiative recombination pathways.

The kinetic role of $Ga^{3+}$ becomes evident when the temporal evolution of the emission linewidth is examined under continued reaction or thermal exposure. In the absence of Ga, CIZS QDs exhibit a gradual linewidth broadening with reaction time (Fig. 2c and Supplementary Fig. 13), consistent with progressive defect regeneration[25]. In contrast, the linewidth of CIZGS QDs remains essentially invariant under identical conditions, indicating that Ga incorporation kinetically locks the low-defect configuration established by $Zn^{2+}$ alloying. This distinction is further amplified during epitaxial ZnS shell growth. When a ZnS shell is grown on CIZS cores, a pronounced low-energy defect band re-emerges, yielding a broadened spectrum characteristic of trap-mediated recombination (Fig. 2d). By contrast, shell growth on CIZGS cores preserves a clean, narrow excitonic emission profile, evidencing suppression of defect regeneration during this processing step.

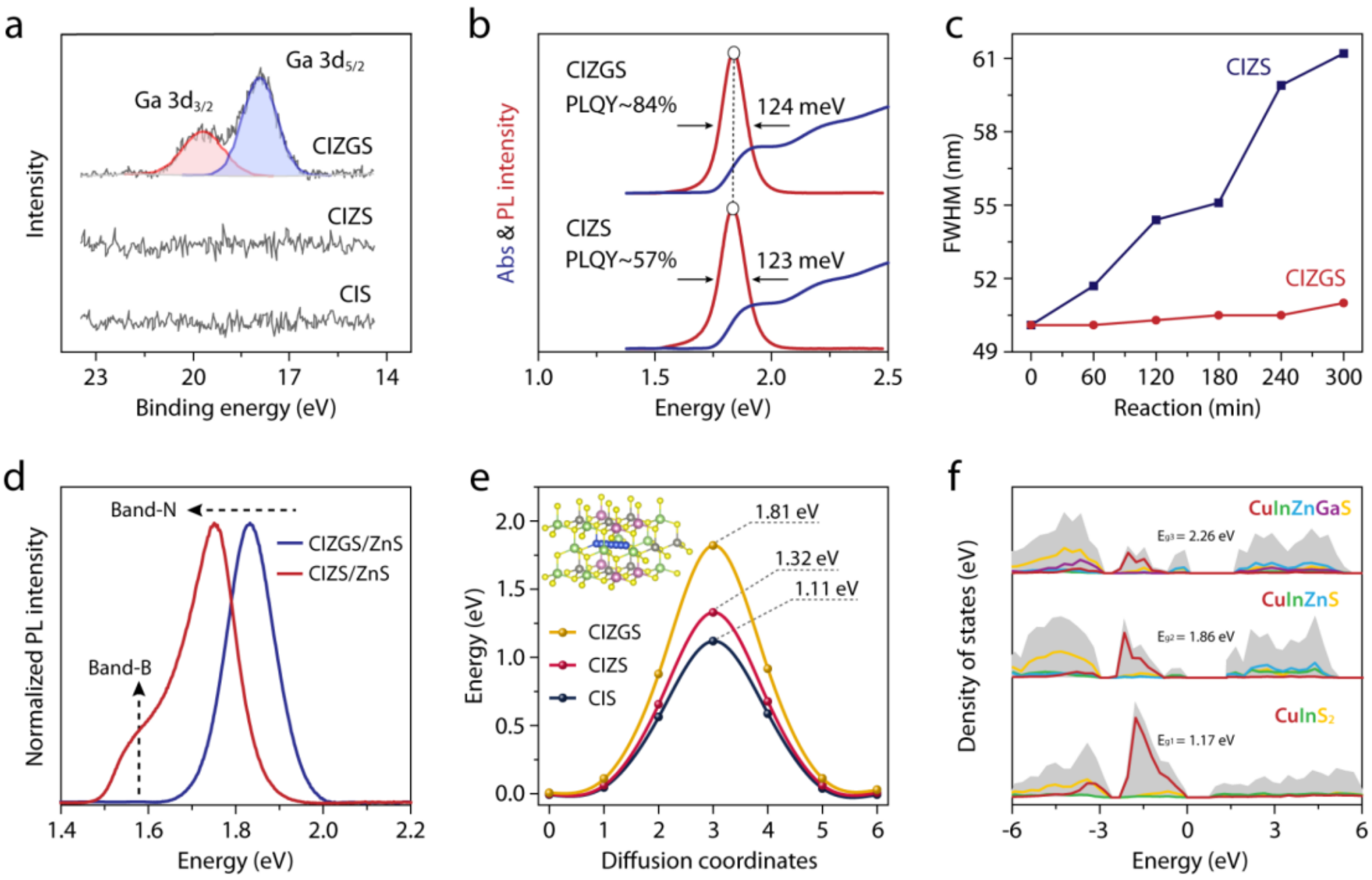

**Figure 2. $Ga^{3+}$ acts as a kinetic "locking" agent that stabilizes the Zn-optimized cation configuration, preserving linewidth while boosting PLQY and suppressing defect regeneration during processing. (a)** High-resolution XPS spectra of Ga 3d orbitals for CIS, CIZS and CIZGS QDs. **(b)** Absorption and PL spectra of CIZS and CIZGS QDs. **(c)** FWHM evolution of CIZS and CIZGS QDs in different heating time at 230 °C under $N_2$ protection. **(d)** Normalized PL spectra of CIZS and CIZGS QDs after the same overgrowth of ZnS shell conditions. **(e)** Variation of the migration distance and energy barrier for $Cu^+$ ions within the lattice of the three types of QDs computed along a representative diffusion pathway. **(f)** DOS plots for the three QDs. The total DOS is shown in grey, with contributions from $Cu^+$ in red, $In^{3+}$ in green, $Zn^{2+}$ in blue, $Ga^{3+}$ in purple, and $S^{2-}$ in yellow.

First-principles calculations provide an atomic-scale rationale for this kinetic stabilization. Energy profiles computed along representative $Cu^+$ migration coordinates show a systematic increase in the energetic penalty for cation displacement upon Ga incorporation, following the trend CIS < CIZS < CIZGS (Fig. 2e). Although the absolute values depend on the specific migration pathway and computational constraints, the monotonic increase captures the experimentally observed resistance to defect reconfiguration in Ga-containing QDs. The density of states (DOS) analysis further reveals the deeper contribution of $Ga^{3+}$ (Fig. 2f). The Ga 3d orbitals introduce new characteristic peaks in the energy bands and form strong hybridization with the S 3p orbitals, resulting in energetically more stable Ga–S bonds. This not only further widens the bandgap to 2.26 eV (Supplementary Fig. 14) but also substantially removes the residual defect states within the forbidden band, leading to a clean semiconductor-like density of states distribution in CIZGS. These results establish $Ga^{3+}$ as a kinetic pinning agent that suppresses $Cu^{+}$ migration and defect regeneration under processing-relevant conditions. By arresting the dynamic instability inherent to the CIS lattice, $Ga^{3+}$ incorporation converts the thermodynamically improved but metastable defect landscape created by $Zn^{2+}$ alloying into a kinetically robust state, enabling durable access to narrow, exciton-like emission.

**Durable excitonic emission during and after shell growth**

Having established a thermodynamically improved and kinetically stabilized core lattice through the combined action of $Zn^{2+}$ alloying and $Ga^{3+}$ pinning, we next evaluate whether this optimized defect landscape can withstand epitaxial shell growth. ZnS passivation is essential for achieving high PLQY and environmental stability in CIS-based QDs, yet it is also a common trigger for defect regeneration due to the elevated temperatures and chemical driving forces involved. In this context, shell growth serves as a stringent stress test for the durability of the excitonic emission state.

Structural characterization confirms the formation of CIZGS/ZnS core/shell QDs. High-angle annular dark-field STEM and elemental mapping show uniform distributions of Cu, In, Zn, Ga, and S without detectable phase segregation (Fig. 3a), indicating coherent shell overgrowth on the stabilized core. Absorption and emission spectra of the resulting QDs retain a narrow excitonic emission band (FWHM ≈ 128 meV), with minimal spectral overlap (Fig. 3b), underscoring that shell growth does not reintroduce defect-mediated broadening.

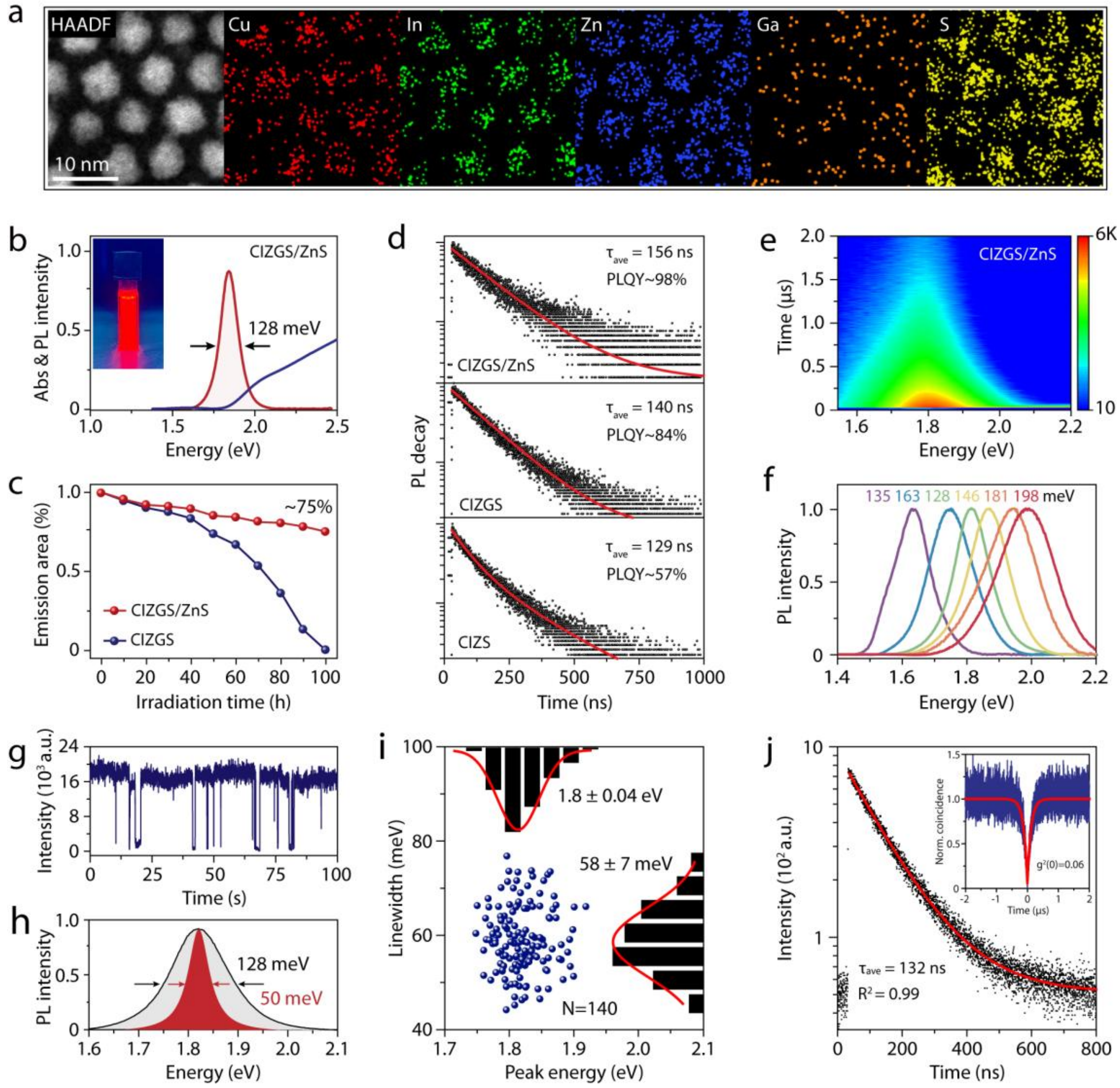

**Figure 3. The thermodynamically narrowed and kinetically locked cores withstand ZnS shell growth, enabling near-unity PLQY with durable excitonic emission and suppressed blinking at the single-dot level. (a)** HAADF-STEM image and corresponding elemental maps of representative CIZGS/ZnS core/shell QDs. **(b)** Absorption and PL spectra of CIZGS/ZnS core/shell QDs. Inset is the digital images of the QDs dispersed in hexane under 405 nm UV-light illumination. **(c)** Evolution of the integrated PL intensity of CIZGS and CIZGS/ZnS QDs under Xe lamp illumination (100 mW cm$^{-2}$). **(d)** PL decay curves of CIZS, CIZGS, and CIZGS/ZnS QDs. The CIZS decay curve was fitted with a biexponential function, whereas those of CIZGS and CIZGS/ZnS were fitted with a mono-exponential function. **(e)** Time-resolved emission spectrum of CIZGS/ZnS core/shell QDs. **(f)** A series of PL spectra of CIZGS/ZnS core/shell QDs by tuning the CIS core sizes. **(g)** Time trace of a single CIZGS/ZnS core/shell QD. **(h)** PL spectra of ensemble (gray shadow) and single (red shadow) CIZGS/ZnS QDs dispersed in hexane and PMMA film, respectively. **(i)** Peak energy and linewidth distribution of single CIZGS/ZnS QDs. The histograms were fitted by Gaussian function. **(j)** PL decay curve of a single CIZGS/ZnS QD, which was fitted with a mono-exponential function. The inset in the upper right corner shows the second-order photon correlation function $g^2(\tau)$ measured using an HBT setup under continuous-wave excitation.

ZnS passivation markedly enhances both radiative efficiency and photostability. The PLQY increases to ~98%, while continuous illumination tests show that CIZGS/ZnS QDs retain ~75% of their initial PL intensity after 100 h (Fig. 3c). Time-resolved PL measurements further reveal a stepwise increase in the average exciton lifetime from CIZS (129 ns) to CIZGS (140 ns) and finally to CIZGS/ZnS (156 ns) (Fig. 3d), in good agreement with progressive suppression of nonradiative decay pathways through kinetic stabilization of the core followed by surface passivation[37]. Such long lifetimes remain compatible with excitonic emission in $CuInS_2$-derived lattices, where weak oscillator strength and exciton fine structure can intrinsically slow radiative recombination at room temperature. Prior theory predicts that the splitting between optically passive and active exciton states can exceed kBT (≈ 26 meV at 300 K), yielding a large Stokes shift together with a long radiative lifetime even for band-edge PL[38]. Consistently, time-resolved emission spectrum shows uniform decay kinetics across the emission band with no long-lived trap-assisted components (Fig. 3e). Importantly, this stabilized excitonic emission remains highly tunable. By controlling the CIS core size (see Supplementary Method 3 for size determination), the PL peak can be continuously shifted from ~620 to 770 nm while maintaining a symmetric, narrow-band lineshape (Fig. 3f).

The intrinsic nature of the stabilized excitonic emission is most convincingly demonstrated at the single-particle level. Individual CIZGS/ZnS QDs exhibit strongly suppressed intensity blinking, remaining in the emissive “on” state for over 80% of the observation time (Fig. 3g). Their emission spectra are well described by a single Lorentzian function with homogeneous linewidths centered around ~58 meV (Fig. 3h and Fig. 3i), significantly narrower than the ensemble linewidth and indicative of band-edge excitonic recombination within individual nanocrystals[39]. Statistical analysis across more than 140 single dots confirms the reproducibility of this narrow linewidth distribution (Fig. 3i and Supplementary Fig. 15). The distribution of single-dot PL peak positions spans the full envelope of the ensemble emission profile (Supplementary Fig. 16), indicating that ensemble broadening arises predominantly from inhomogeneity in dot size and composition. Second-order photon correlation measurements yield $g^2(0)$ values as low as 0.06 (inset in Fig. 3j), confirming high-purity single-photon emission at room temperature and verifying that the following measurements reflect true single-dot photophysics[40,41]. Statistical analysis of the lifetime of individual CIZGS/ZnS QDs reveals a distribution tightly clustered around 136 ns (Fig. 3j, Supplementary Fig. 17 and 18), further confirming the homogeneity of exciton recombination pathways within the nanocrystal ensemble.

These ensemble- and single-particle observations demonstrate that ZnS shell growth, when applied to a thermodynamically and kinetically stabilized CIZGS core, functions primarily as a surface passivation step rather than a defect-activation process. The resulting core/shell QDs therefore sustain narrow, efficient, and photostable excitonic emission under processing conditions that typically compromise CIS-based emitters, completing the transition from a defect-dominated to an intrinsically excitonic recombination regime.

To benchmark our results against the state of the art, we compared key parameters of CIZGS/ZnS QDs with recently reported high-performance cadmium-free QDs (Supplementary Table 4). The thermodynamic–kinetic decoupling strategy proposed in this work enables the FWHM of $CuInS_2$-based QDs to be sharpened to approximately 120 meV. This achievement ranks among the leading performance levels for cadmium-free QDs in the deep-red to near-infrared spectral region. Notably, while delivering exceptional color purity, the strategy also achieves a PLQY approaching unity and demonstrates outstanding photostability. Thus, it successfully addresses the long-standing challenge of simultaneously optimizing high color purity, high quantum efficiency, and high stability.

**Implications for device-relevant optical performance**

The availability of narrow, stable excitonic emission in CIZGS/ZnS QDs has direct implications for optoelectronic architectures in which spectral purity governs device physics rather than merely aesthetic color quality. Luminescent solar concentrators (LSCs) provide a particularly sensitive platform to evaluate this effect, as their optical efficiency is fundamentally limited by reabsorption losses arising from spectral overlap between absorption and emission[42,43]. In this context, linewidth narrowing offers a materials-level route to mitigate a dominant loss mechanism without altering device geometry.

Fig. 4a illustrates the LSC configuration employed in this study, in which CIZGS/ZnS QDs are embedded in a UV-curable polymer matrix and coupled to a silicon solar cell at the waveguide edge. The absorption and emission spectra of the QDs (Fig. 4b) exhibit a narrow excitonic emission band with minimal overlap with the absorption edge, a spectral characteristic that is rarely achieved in CIS-based emitters. Photographs of the resulting LSC plates under ambient and UV illumination further highlight the combination of high visible transparency and intense, spatially uniform red emission enabled by this narrow-band response.

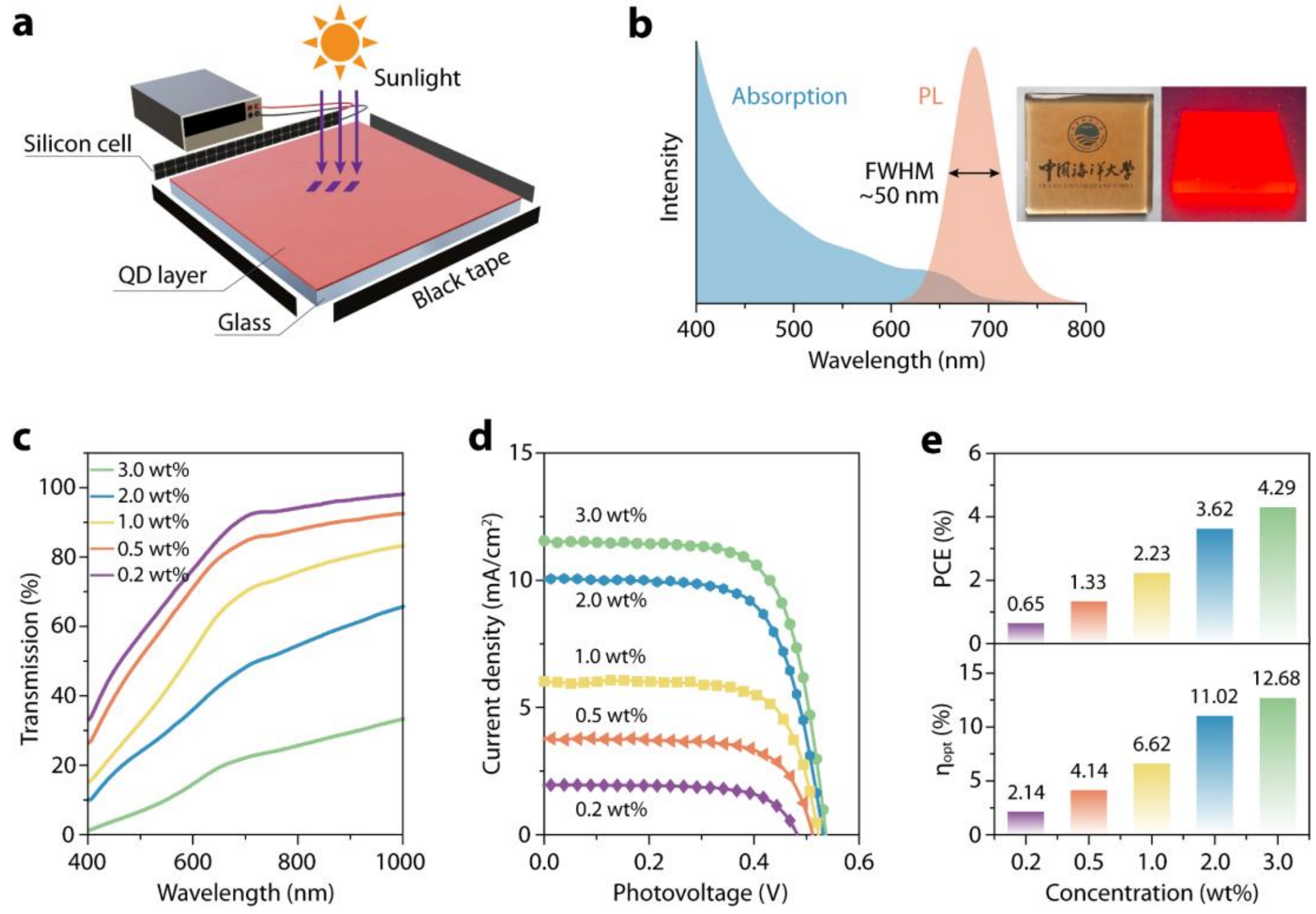

**Figure 4. Narrow, stable excitonic emission translates into reduced reabsorption losses and improved optical efficiency in luminescent solar concentrators. (a)** Diagram of QD-based LSC devices. **(b)** Absorption and PL spectra of typical CIZGS/ZnS QDs used for LSC fabrication. Insets are the photographs of LSC waveguide (5 cm × 5 cm × 0.5 cm). **(c)** Transmission spectra of LSCs with different QD doping concentrations. **(d)** *J–V* characteristics of the devices with different QD doping concentrations. **(e)** Corresponding statistical summaries of power conversion efficiency (PCE) and external optical efficiency ($\eta_{opt}$).

As the QD loading is increased, the optical transparency of the LSC decreases in a predictable manner (Fig. 4c and Supplementary Fig. 19), while the edge-collected photocurrent rises monotonically. Meanwhile, as the QD concentration increases, the PL emission peak shows a red shift of approximately 1–8 nm (Supplementary Fig. 20), which is speculated to arise from the enhanced self-absorption effect and/or unavoidable aggregation of QDs at high concentrations during the solidification process. Current–voltage measurements reveal a systematic increase in short-circuit current density ($J_{sc}$) with increasing QD concentration (Fig. 4d), reflecting more efficient harvesting and waveguiding of the emitted photons (the J–V curve data for the silicon cell are shown in Supplementary Fig. 21). Correspondingly, the external optical efficiency ($\eta_{opt}$) increases to a maximum value of 12.68% at a loading of 3.0 wt% (Fig. 4e).

The detailed photovoltaic parameters are summarized in Supplementary Table 5 (see Supplementary Method 4 for calculation details). Notably, the performance improvement cannot be attributed solely to enhanced absorption; rather, it reflects the reduced probability of photon reabsorption during propagation enabled by the intrinsically narrow emission of the QDs. For an LSC with dimensions of 5×5×0.5 $cm^3$, the PL intensity of the QDs decreases as the distance between the excitation spot and the device edge increases, eventually stabilizing beyond a distance of 3–5 cm (Supplementary Fig. 22). This trend is consistent with the mechanisms of reabsorption loss and scattering loss during waveguide propagation, and aligns with previous reports of distance-dependent performance degradation in LSCs.[44,45]

Regarding device stability, we performed continuous light-soaking tests under standard xenon lamp illumination (100 $mW/cm^2$) on encapsulated LSCs. The prototype devices exhibit a gradual efficiency decay, retaining ~40% of the initial value after 50 h (Supplementary Fig. 23). Given that the same CIZGS/ZnS QDs preserve ~75% of their PL intensity after 100 h in solution (Fig. 3c), the device-level decay is likely dominated by matrix- and processing-related factors (e.g., polymer photochemistry and dispersion evolution during curing) rather than by an intrinsic loss of excitonic emissivity from the core/shell QDs. Further optimization of the host matrix and curing chemistry should therefore provide a direct route to improved operational stability.

To evaluate the performance level of the LSCs prepared in this work, we systematically compared their optical efficiency with recently reported cadmium-free QD-based LSCs (Supplementary Table 6). While maintaining good optical transparency, the present device achieves an optical efficiency of 12.68%, surpassing many reported counterparts and being comparable to leading cadmium-free QD-based LSCs. This result demonstrates that the CIZGS/ZnS QDs prepared based on the thermodynamic-kinetic decoupling strategy not only possess excellent intrinsic optical properties but also exhibit significant application potential in practical device integration, offering a competitive material solution for the development of efficient and eco-friendly luminescent concentrator technology.

These results underscore that the spectral narrowing achieved through thermodynamic–kinetic defect control translates directly into a device-relevant advantage. While the present LSCs are not optimized for long-term operational stability or absolute efficiency records, they serve to demonstrate a key physical consequence of suppressing defect-mediated broadening in CIS-based QDs. By reducing reabsorption losses at the materials level, stable excitonic emission allows LSCs to operate closer to their intrinsic optical limits, illustrating how defect-landscape engineering can unlock performance gains in architectures that are otherwise constrained by the spectral characteristics of the luminophore.

## Discussion

The long-standing difficulty in achieving narrow-band, stable emission in defect-tolerant multinary QDs is often framed as an incomplete passivation problem. Our results suggest a more fundamental interpretation. In CIS-based nanocrystals, the persistence of broad defect-mediated PL, even as PLQYs approach unity, reflects not only the low thermodynamic cost of forming native defects, but also the kinetic accessibility of pathways that regenerate these defects under processing-relevant perturbations. In other words, the defect landscape in such "soft" semiconductors is defined by both an energetically favored basin (defect formation) and a dynamically accessible network of return routes (defect regeneration), the latter being enabled by the unusually high mobility of $Cu^+$ in the cation sublattice.

By explicitly decoupling these two historically intertwined contributions, the present "strain-and-pin" strategy establishes a thermodynamic–kinetic route to durable band-edge recombination. $Zn^{2+}$ alloying imposes a lattice-level perturbation evidenced by the emergence of a measurable compressive strain field, and correspondingly raises the calculated formation energy of the dominant $V_{Cu}$, providing a thermodynamic basis for suppressing defect populations and narrowing the emission toward an exciton band. However, the partial recovery of defect signatures upon heating underscores that this improved state is not automatically durable. Rather, it remains kinetically metastable in an ion-mobile lattice. This explains why strategies that focus primarily on defect formation energetics, composition tuning, shell passivation, or conventional alloying, can yield high efficiencies yet still fail to deliver persistent spectral purity in Cu-based I–III–VI systems.

The critical role of kinetic stabilization is revealed by $Ga^{3+}$ incorporation. Ga does not primarily function as an additional lever to further narrow the emission. Instead, it locks the thermodynamically optimized defect landscape against $Cu^+$ migration. Experimentally, this is manifested in the invariant linewidth of CIZGS under continued reaction conditions and, most decisively, in the preservation of a clean narrow-band spectrum after ZnS shell growth, in contrast to the defect-band re-emergence observed in the Ga-free control. These results indicate that the decisive failure mode for narrow-band emission in CIS-based QDs is not merely the presence of defects, but the ease with which they can be regenerated during processing. From an atomistic viewpoint, Ga–S bonding introduces robust pinning motifs that increase the energetic penalty for $Cu^{+}$ displacement along representative diffusion coordinates, thereby reducing the kinetic connectivity between the low-defect excitonic state and defect-rich configurations. The outcome is a defect landscape that is both thermodynamically shifted and kinetically insulated.

The durable nature of the resulting excitonic state is corroborated at multiple levels of stringency. Ensemble spectroscopy shows that ZnS overgrowth on kinetically stabilized cores functions purely as surface passivation rather than a defect-activation step, enabling near-unity PLQY without reintroducing low-energy defect bands. Time-resolved emission mapping further supports a uniformly excitonic recombination channel across the entire emission band, with no evidence of slow trap-assisted components. Most importantly, single-particle measurements show suppressed blinking, homogeneous Lorentzian linewidths, and high-purity single-photon emission at room temperature. These signatures jointly rule out ensemble averaging, size-selection artifacts, or heterogeneous local environments as the origin of the observed spectral narrowing, and instead establish that band-edge recombination has been stabilized within individual nanocrystals.

Beyond this specific material system, the broader implication is a general design principle for defect-tolerant and ion-mobile semiconductors. Achieving intrinsic optoelectronic properties requires engineering not only the thermodynamic tendency to form defects but also the kinetic accessibility of defect-regeneration pathways. The “strain” component in this framework provides a route to reshape defect formation energetics through lattice-level perturbations, whereas the "pinning" component identifies and immobilizes the dominant mobile species by introducing strong-bonding motifs within the relevant sublattice. This logic is expected to extend to other multinary chalcogenides in which cation disorder and ion migration undermine spectral purity and stability. For example, in Ag-based I–III–VI nanocrystals, where $Ag^{+}$ mobility and antisite disorder are similarly prevalent, targeted incorporation of elements that form strong bonds with the anion framework may serve as kinetic pinning agents. More broadly, the same thermodynamic–kinetic partitioning is conceptually aligned with stabilization strategies in other ion-mobile semiconductors, including lead-free halide perovskites[46,47], where defect formation and ion migration together govern long-term optical and electronic stability.

Finally, the device-level consequence demonstrated in LSCs highlights why spectral purity is not merely a cosmetic metric but a fundamental physical lever. In architectures constrained by reabsorption, linewidth narrowing directly reduces self-absorption losses during photon transport, enabling higher optical efficiency without relying on geometric or optical design complexity. In this sense, thermodynamic–kinetic defect control provides a materials-level route to unlock performance gains in spectral-overlap-limited devices. Collectively, our results show that the persistent "efficiency–purity–stability" compromise in defect-tolerant QDs is not inevitable. It can be overcome by a coordinated approach that decouples and simultaneously controls defect formation energetics and defect regeneration kinetics.

## Conclusions

We have established a thermodynamic–kinetic "strain-and-pin" strategy that renders excitonic emission durable in defect-tolerant $CuInS_2$ quantum dots. $Zn^{2+}$ alloying raises the formation energy of native copper vacancies, narrowing the emission to an excitonic line. $Ga^{3+}$ incorporation suppresses $Cu^{+}$ migration, locking in this state during ZnS shell growth. The resulting Cd-free core/shell dots achieve near-unity quantum yield, homogeneous single-dot linewidths down to ~58 meV, suppressed blinking, and room-temperature single-photon purity ($g^2(0) = 0.06$), a clear evidence that band-edge exciton recombination is stabilized at the single-particle level. More broadly, this work shows that in ion-mobile, defect-prone semiconductors, intrinsic optoelectronic properties become accessible only when defect formation is thermodynamically suppressed and defect regeneration kinetically blocked. The device relevance is underscored in luminescent solar concentrators, where the stabilized narrowband emission reduces reabsorption losses, yielding an external optical efficiency of 12.68%. Our findings thus break the longstanding efficiency–purity–stability compromise in eco-friendlier quantum emitters, opening a viable path toward heavy-metal-free sources for quantum optics and photonics.

## Associated Content

### Supporting Information

Experimental details; Characterization methods; Supplementary methods for PLQY measurements, DFT calculations, size calculations and LSC performance metrics; PL, PLE, TRPL spectra of various CIS-based QDs; TEM, GAP, XRD, XPS and ICP analysis of CIS-based QDs; PL spectra, decay curves and corresponding statistics of individual CIZGS/ZnS QDs; LSC device characterization and stability of CIS-based QDs together with benchmarks of heavy-metal-free narrow-emitting QDs.

## Author Information

### Corresponding Author

Lei HOU − Engineering Research Center of Optical Instrument and System, Ministry of Education and Shanghai Key Lab of Modern Optical System, University of Shanghai for Science and Technology, Shanghai 200093, China

Email: h_lei@pku.edu.cn

Lixin CAO − School of Materials Science and Engineering, Ocean University of China, No. 1299, Sansha Road, Huangdao District, Qingdao, 266404, China

Email: caolixin@ouc.edu.cn

Chenghui Xia − School of Materials Science and Engineering, Ocean University of China, No. 1299, Sansha Road, Huangdao District, Qingdao, 266000, China

Email: c.xia@ouc.edu.cn; ORCID: 0000-0001-5087-8805

## Competing Interests

The authors declare no competing interests.

## Acknowledgments

C.X. acknowledges the financial support from TaiShan Scholar Foundation of Shandong Province (Grant No.: tsqn202306113), Natural Science Foundation of Shandong Province (Grant No.: 2024HWYQ-037, ZR2023QB174), Qingdao Postdoctoral Science Foundation (Grant No.: QDBSH20230201024). X.X. acknowledges the funding support from Fundamental Research Funds for the Central Universities (No. 21008044C2001). L.H. acknowledges financial support from the National Natural Science Foundation of China (Grant No.: 12474425) and Natural Science Foundation of Shanghai under the 2024 Shanghai Action Plan for Science, Technology and Innovation (Grant No.: 24ZR1453000).

# Supplementary Information:

## Materials, Synthesis, and Characterization

**Materials.** Copper(I) acetate [CuAc, 97%], trioctylphosphine oxide (TOPO, 98%), 1-octadecene (ODE, 90%), 1-dodecanethiol (DDT, 98%), anhydrous indium acetate [$In(Ac)_3$, 99.99%], trioctylphosphine (TOP, 90%), n-octylamine (OctAm, 99%), zinc iodide ($ZnI_2$, 99.99%), anhydrous zinc acetate [$Zn(Ac)_2$, 99%], oleic acid (OA, 90%), anhydrous gallium(III) chloride [$Ga(Cl)_3$, 98%], diphenylphosphine (DPP, 95%), anhydrous cyclohexane, toluene, methanol, and butanol were purchased from Aladdin®. TOPO, ODE, and OA were separately degassed under vacuum at 120 °C prior to use in the synthesis.

**Synthesis of $Cu_{2-x}S$ NCs.** $Cu_{2-x}S$ nanocrystals were synthesized through a high-temperature organometallic route. Briefly, copper acetate, TOPO, and ODE were mixed and degassed under vacuum at elevated temperature to remove residual moisture and oxygen. The mixture was then heated under nitrogen, followed by injection of DDT at an intermediate temperature. The reaction was maintained at high temperature for a controlled period to tune the nanocrystal size. After cooling to room temperature, the crude product was purified by repeated precipitation–centrifugation cycles using mixed polar solvents and finally redispersed in cyclohexane under an inert atmosphere for storage.

**Synthesis of CIS NCs.** Wurtzite $CuInS_2$ nanocrystals were obtained by a template-assisted cation-exchange reaction using $Cu_{2-x}S$ NCs as the starting template. An indium precursor was first prepared by dissolving an indium salt with TOP in ODE under nitrogen, followed by vacuum degassing. The purified $Cu_{2-x}S$ NCs were redispersed in a mixed coordinating solvent containing DDT and ODE and subsequently introduced into the indium precursor solution. The cation-exchange reaction was carried out at moderate temperature for several hours. The resulting wz-CIS NCs were purified by repeated precipitation and centrifugation, redispersed in cyclohexane, and stored in a glovebox. When aggregation or gelation occurred during purification, a small amount of alkylamine was added to assist redispersion.

**Synthesis of CIZS NCs by Zn Alloying.** CIZS nanocrystals were prepared by post-synthetic Zn incorporation into wz-CIS NCs. A purified CIS NC dispersion was mixed with ODE and degassed under vacuum. After switching to nitrogen, a zinc precursor solution containing a zinc halide, TOP, and ODE was rapidly injected. The reaction temperature was then increased and maintained for a defined period to promote Zn incorporation and lattice alloying. After completion, the reaction mixture was cooled naturally and stored under an inert atmosphere.

**Ga Incorporation into CIZS NCs.** Ga-modified CIZS nanocrystals were prepared through a mild cation-exchange process. A gallium precursor was prepared in a nitrogen-filled glovebox by reacting an anhydrous gallium salt with an organophosphorus ligand, followed by dilution with anhydrous solvent. A small aliquot of this precursor was introduced into the as-prepared CIZS NC solution, and the reaction was conducted at relatively low temperature for a controlled time. The product was isolated by precipitation with mixed alcohol solvents, centrifuged, and redispersed in cyclohexane.

**Synthesis of CIZGS/ZnS QDs by Shell Coating.** ZnS shell growth was performed using a pre-prepared zinc–sulfur precursor. The shell precursor was obtained by dissolving a zinc salt in a mixture of fatty acid and thiol ligands under vacuum and heating until a clear solution was formed. For shell coating, purified CIZGS NCs were dispersed in a mixed solvent containing DDT and ODE, degassed, and heated under nitrogen. After temperature stabilization, the ZnS precursor was rapidly injected, and the reaction was maintained at elevated temperature to allow shell growth. The crude CIZGS/ZnS QDs were purified by repeated precipitation–centrifugation cycles and finally dispersed in cyclohexane for storage under an inert atmosphere.

**Fabrication of LSC Devices.** A predetermined amount of CIZGS/ZnS quantum dots (QDs) was mixed with toluene, followed by the addition of 3 mL UV-curable resin (65% polyurethane acrylate oligomer, 20% monomer, 10% photoinitiator, 5% defoaming agent). The mixture was stirred to form a homogeneous slurry, then subjected to ultrasonication for degassing. The resulting suspension was coated onto a glass substrate ($5\times5\times0.5$ $cm^3$, low-iron glass) via doctor-blading and exposed to UV light (365 nm) for 10 s to fabricate the LSC device.

**Characterization.** Samples for optical testing were prepared by dispersing CIZGS/ZnS QDs in cyclohexane within a 1 cm path length quartz cuvette inside a glovebox. Absorption spectra were acquired using a Metash UV-8000 spectrophotometer. PL spectra were collected on an Edinburgh FLS 980 spectrometer with 442 nm excitation. Absolute photoluminescence quantum yield calculations are detailed in Supporting Method 1. Time-resolved PL measurements were performed using a 450 nm laser source. For single-dot spectroscopy, QDs were dispersed in toluene (xenon lamp-treated), diluted in a 3 wt% poly(methyl methacrylate) solution, and spin-coated onto glass slides. Individual QDs were imaged using a scanning confocal microscope coupled with a HORIBA iHR550 spectrometer under 532 nm laser excitation. XRD patterns were obtained on a Bruker D8 Advance diffractometer with a Cu Kα X-ray source ($\lambda$ = 1.5406 Å) using QD films coated on quartz substrates. HR-TEM and HAADF-STEM imaging were performed on a Thermo Fisher Spectra 300 S/TEM operated at 300 kV (camera length: 115 mm), equipped with a ChemiSTEM EDS detector for elemental mapping. XPS measurements were conducted on a Thermo Scientific ESCALAB Xi+ system with a monochromatic Al Kα source (E = 1486.68 eV) at 14.8 kV and 10.8 mA. Survey scans used a 100 eV pass energy, while high-resolution scans used 20 eV. Binding energies were calibrated against the C 1s peak at 284.8 eV. ICP-OES analysis was performed on an Agilent 5800 instrument after digesting dried samples in aqua regia and diluting with $HNO_3$ to ppm concentrations. LSC device characterization methods are described in Supplementary Method 4.

## Supplementary Methods

### Supplementary Method 1: Details of Fluorescence Quantum Yield Measurement and Calculation

The photoluminescence quantum yield (PLQY), denoted as $\boldsymbol{\eta}$, is defined as the ratio of the total number of emitted photons to the total number of absorbed photons. The sample preparation for PLQY measurement is as follows: QDs are diluted in cyclohexane such that the absorbance at the excitation wavelength of 442 nm is below 0.1. The diluted solution is then transferred into a 10 mm path length quartz cuvette. The standard sample holder in the sample compartment is replaced with an integrating sphere, and the two associated lens components must be removed.

First, the quartz cuvette containing pure cyclohexane solvent is placed inside the integrating sphere as a reference and excited at 442 nm. To ensure 100% reflectivity, the slit widths should be adjusted so that the scattered signal at 442 nm reaches maximum sensitivity without exceeding the detector's saturation level (e.g., excitation slit = 5; emission slit = 0.2). The reference emission spectrum from 412 to 800 nm is acquired (blue curve in Fig. S1). The prepared sample is then measured under the same conditions (red curve). The number of absorbed photons can be determined from the integrated difference of the two scattering curves at 442 nm, while the number of emitted photons is obtained from the integrated difference of the emission spectra. The PLQY can be calculated using the following equation:

$$\eta = \frac{E_B - E_A}{S_A - S_B}.$$

where $\boldsymbol{E_A}$ and $\boldsymbol{E_B}$ represent the integrated emission values of the reference and the sample, respectively, over the wavelength range of 560 to 800 nm, and $\boldsymbol{S_A}$ and $\boldsymbol{S_B}$ denote the integrated scattering values of the reference and the sample, respectively, over the wavelength range of 412 to 450 nm.

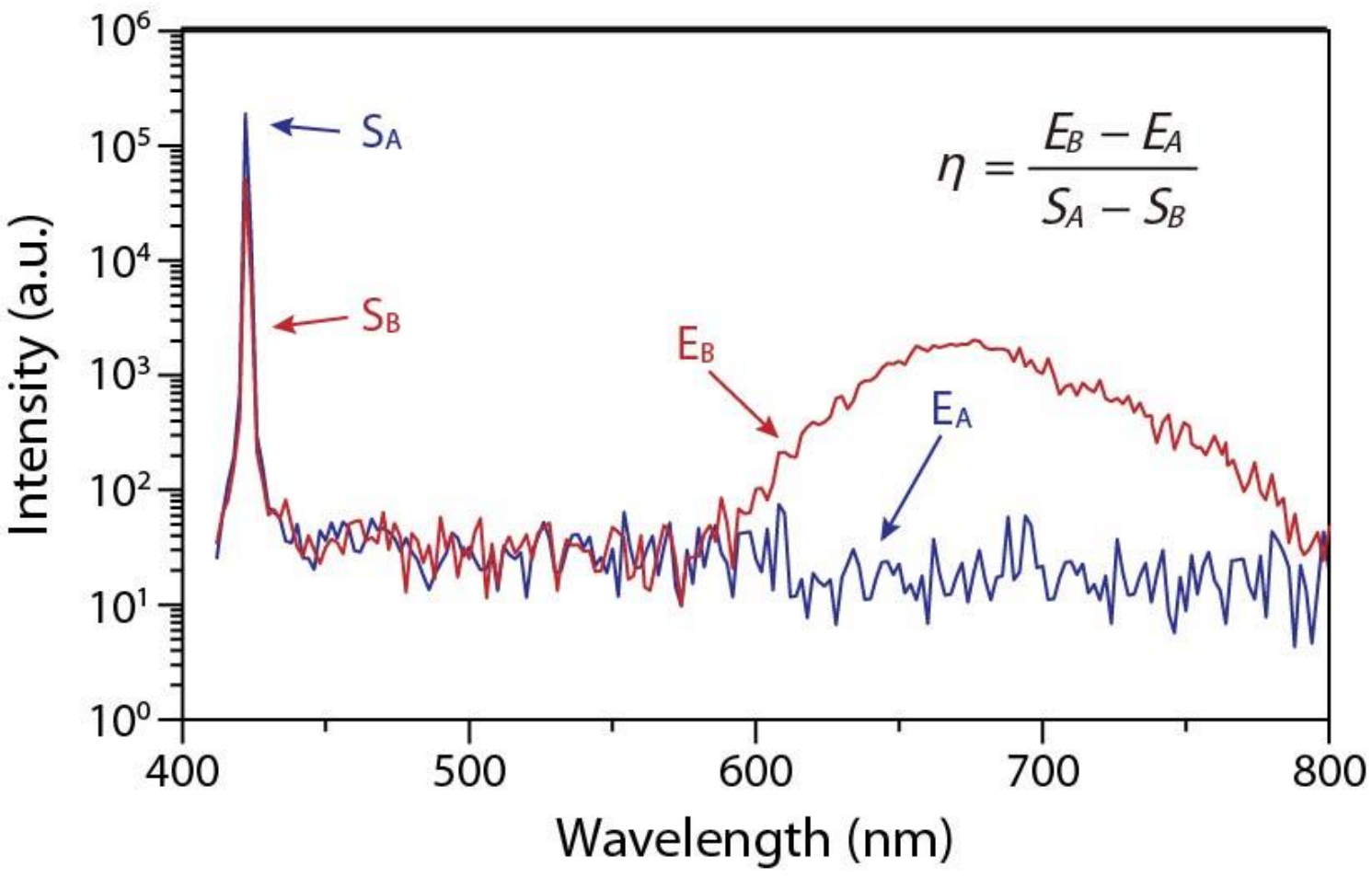

**Figure S1.** PL emission spectra of the reference sample (blue curve) and the test sample (red curve). The excitation wavelength was 442 nm, and the detection wavelength range was set from 412 to 800 nm. The excitation and emission slit widths were fixed at 5 and 0.2, respectively. All measurements were performed under identical conditions.

**Supplementary Method 2: Density Functional Theory (DFT) Calculation Method**

First-principles calculations were performed using the Vienna Ab initio Simulation Package (VASP) [1]. The exchange–correlation interactions were treated within the generalized gradient approximation (GGA) using the Perdew–Burke–Ernzerhof (PBE) functional[2]. The ion–electron interactions were described by the projector augmented-wave (PAW) pseudopotentials[3]. A plane-wave kinetic energy cutoff of 500 eV was employed to ensure the accuracy of the electronic structure calculations.

All structures were constructed based on a three-dimensional periodic bulk supercell model. A 3D supercell was generated by expanding the optimized primitive cell along the lattice vectors to eliminate significant interactions between migrating ions and their periodic images, thereby ensuring that the calculated migration energy barriers accurately reflect the intrinsic ion diffusion behavior in the bulk phase.

Brillouin-zone integrations were carried out using a Γ-centered Monkhorst–Pack k-point mesh of $7 \times 7 \times 7$. Systematic convergence tests with respect to the k-point density and supercell size were performed to guarantee the reliability of the total energies and migration barriers. All atomic positions were fully relaxed under fixed lattice parameters until the total energy and atomic forces converged to $1 \times 10^{-6}$ eV and 0.01 eV $Å^{-1}$, respectively.

Ion migration energy barriers were calculated using the climbing-image nudged elastic band (CI-NEB) method. The initial and final states corresponded to relaxed configurations with the ion located at neighboring stable lattice sites, and five intermediate images (seven images in total) were inserted between them to describe the complete migration pathway. All NEB images were fully relaxed until the maximum force on each atom was less than 0.05 eV $Å^{-1}$.

During the migration energy calculations, the system was maintained under electrically neutral conditions, and different charge states were not explicitly considered. Since a three-dimensional periodic bulk model was adopted and no charged defects were involved, no additional chemical potential corrections or finite-size charge corrections were required. The migration energy barrier was determined from the relative energies of the initial, transition, and final states; therefore, chemical potential terms cancel out and do not affect the physical reliability of the calculated results. To more accurately account for possible long-range interactions, the DFT-D dispersion correction method was employed[4].

### Supplementary Method 3: Size Calculation of wz-CIS QDs

According to the equation reported in previous literature, the average size of the as-synthesized wurtzite CIS QDs was calculated as follows[5]:

$$E_g(wz_CIS) = 1.327 + \frac{1}{0.0125d^2 + 0.225d + 0.938} \quad \text{Eq. S1}$$

where $\boldsymbol{E_g}$ represents the energy (in electron volts, eV) corresponding to the first absorption transition peak of the wurtzite CIS QDs, and $\boldsymbol{d}$ denotes the size. The energy of the first absorption transition was determined by locating the minimum of the second derivative of the absorption spectrum shown in Figure S2.

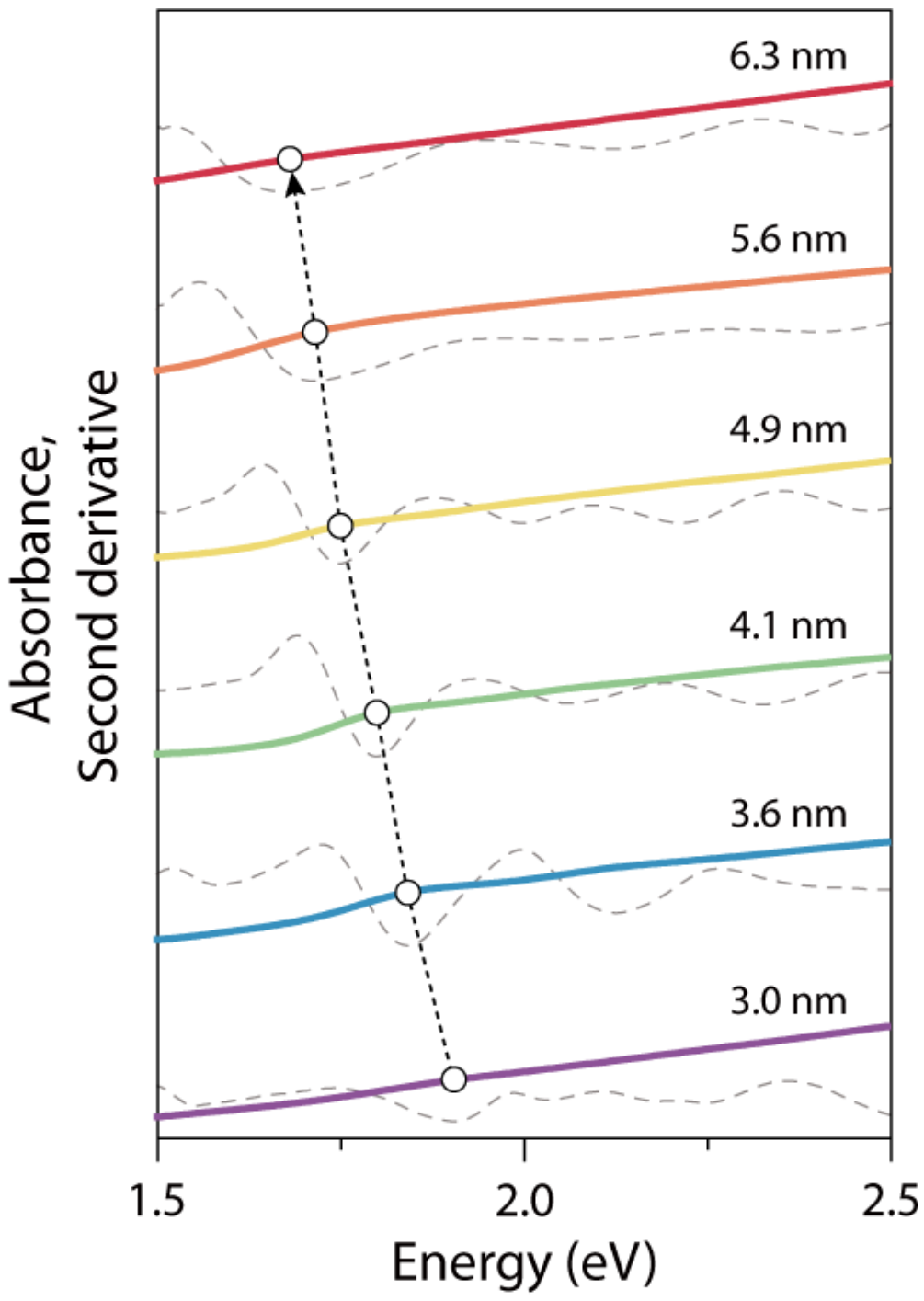

**Figure S2.** Absorption spectra and corresponding second-derivative curves of CIS QDs synthesized from $Cu_{2-x}S$ template nanocrystals with varying sizes. The sizes were controlled by varying the growth time of $Cu_{2-x}S$ nanocrystals (5–50 min). The gray dashed lines represent the second-derivative curves of the corresponding absorption spectra. The first absorption transition energy was determined by locating the minima of the second-derivative curves.

**Table S1.** The first absorption transition energy and corresponding calculated sizes of CIS QDs derived from Eq. S1.

| Sample NO. | Growth time (min) | First Absorption Peak | | Size (nm) |
|---|---|---|---|---|
| | | Wavelength (nm) | Energy (eV) | |
| 1 | 5 | 651 | 1.90 | 3.02 |
| 2 | 10 | 671 | 1.85 | 3.63 |
| 3 | 15 | 685 | 1.81 | 4.10 |
| 4 | 25 | 707 | 1.75 | 4.91 |
| 5 | 35 | 724 | 1.71 | 5.61 |
| 6 | 50 | 740 | 1.68 | 6.34 |

**Supplementary Method 4: Methods for Calculating the Power Conversion Efficiency, Geometric Gain Factor, and Optical Efficiency of LSCs**

The performance of QD-based luminescent solar concentrators (LSCs) was characterized under dark conditions using illumination from a solar simulator. A silicon solar cell was attached to the edge of the LSC. The exposed area of the cell was covered with black tape following a previously reported method to prevent stray light incidence[6]. The LSC was then subjected to I–V characterization under simulated AM 1.5G illumination with perpendicular incidence to its surface.

The power conversion efficiency of the solar concentrator can be calculated using the following formula[7]:

$$PCE = \frac{I_{SC}V_{OC}FF}{P_0 A_{active}} = \frac{J_{SC}V_{OC}FF}{P_0}$$

where $\boldsymbol{I_{SC}}$ is the short-circuit current, $\boldsymbol{J_{SC}}$ is the short-circuit current density, $\boldsymbol{V_{OC}}$ is the open-circuit voltage, $\boldsymbol{FF}$ is the fill factor, $\boldsymbol{P_0}$ is the power density under standard AM 1.5G solar spectrum illumination, and $\boldsymbol{A_{active}}$ is the active illumination area for incident light collection. All parameters are obtained during testing.

The geometric gain factor of the solar concentrator can be calculated using the following formula[8]:

$$G = \frac{A_{active}}{A_{edges}}$$

The geometric gain factor, $\boldsymbol{G}$, is defined as the ratio of the top surface (light-receiving area, $\boldsymbol{A_{active}}$) to the edge area (light-emitting area, $\boldsymbol{A_{edges}}$) of the device, quantifying its geometric ability to concentrate incident light. It represents the theoretical maximum factor by which light collected on the top surface is concentrated onto the edge-mounted photovoltaic cells.

The optical efficiency of the concentrator can be calculated using the following formula[9]:

$$\eta_{opt} = \frac{J_{LSCs}}{J_{SC} \times G}$$

$\boldsymbol{J_{LSCs}}$ and $\boldsymbol{J_{SC}}$ represent the short-circuit current density of the photovoltaic cell when connected to LSCs and the short-circuit current density of the photovoltaic cell under direct illumination, respectively.

**Thermodynamic Suppression of Defect Formation Via $Zn^{2+}$ Alloying**

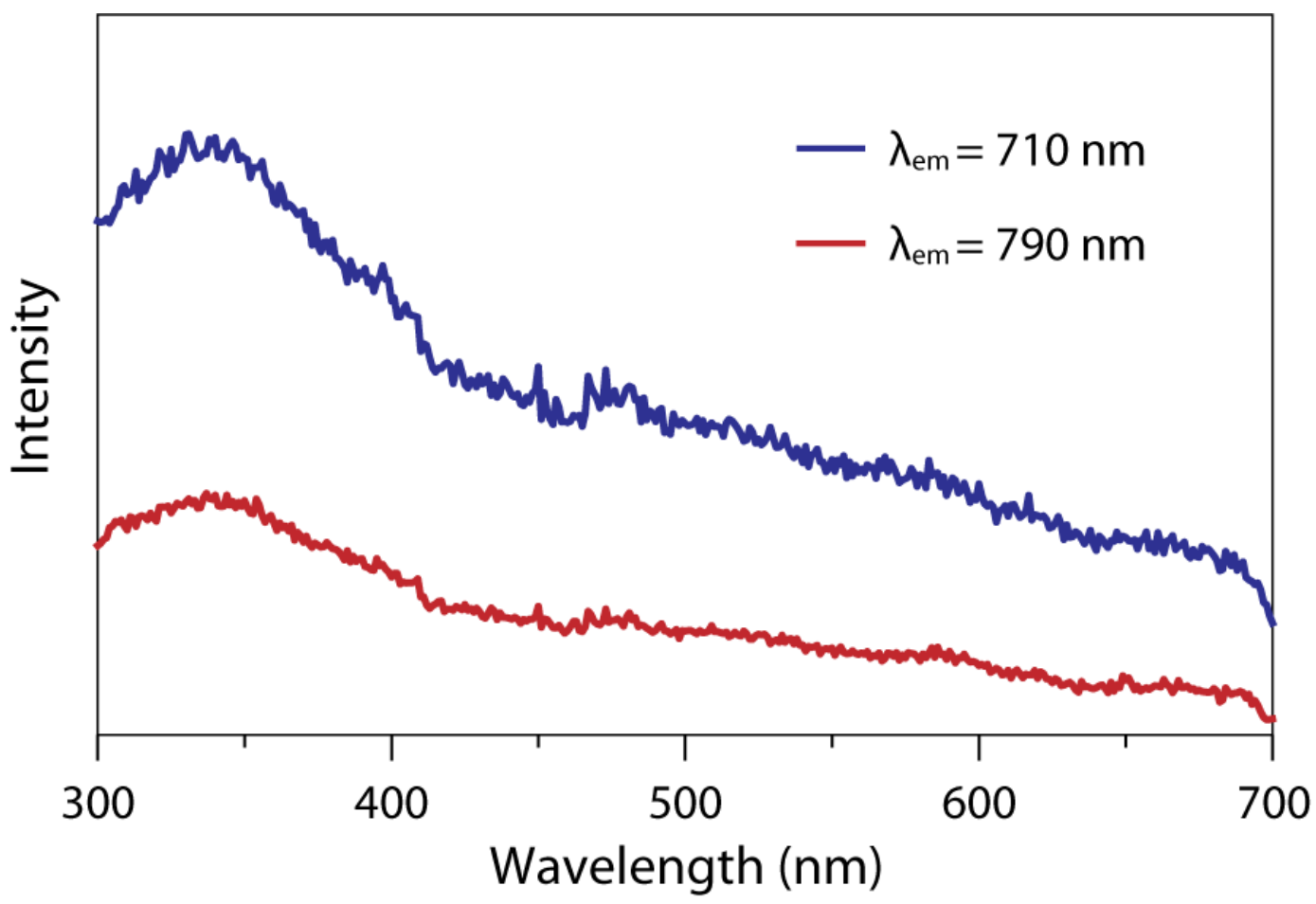

**Figure S3.** Photoluminescence excitation (PLE) spectra of CIZS QDs measured at emission wavelengths of 710 nm and 790 nm.

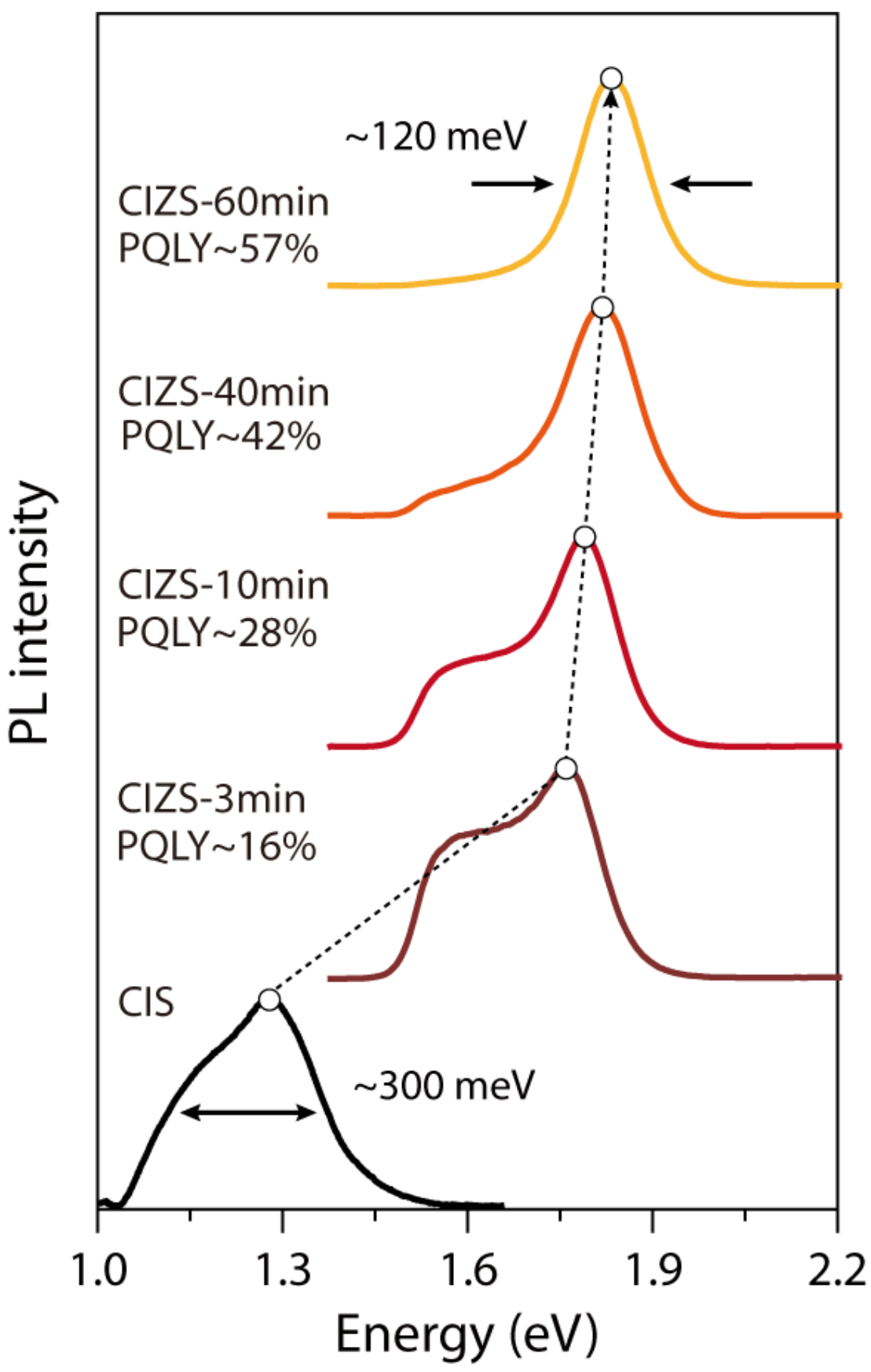

**Figure S4.** Evolution of the PL spectra and PLQY during the extended Zn-alloying reaction.

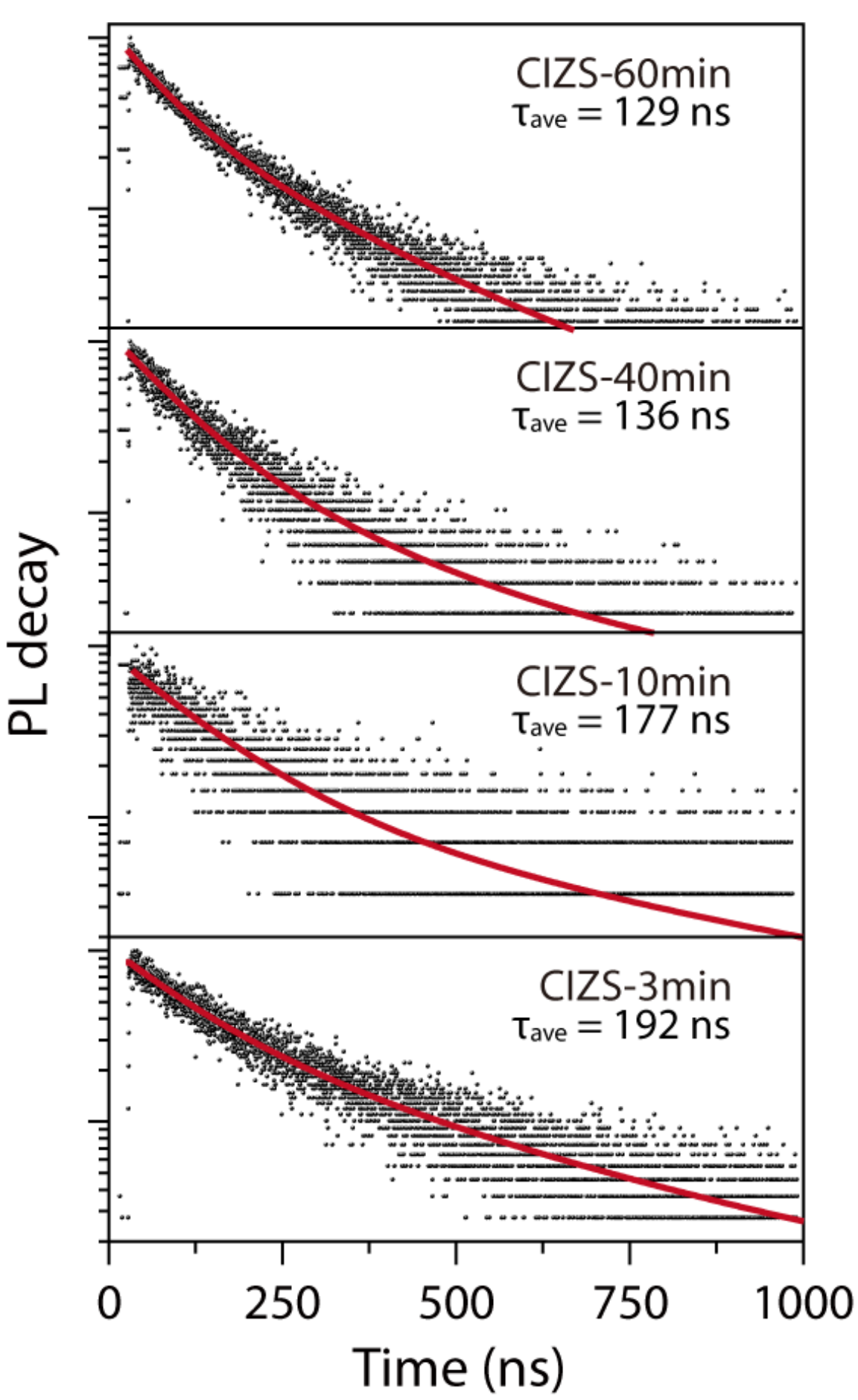

**Figure S5.** PL decay curves (monitored at their respective peak wavelengths) of the CIZS QDs corresponding to the different reaction times in Figure S4. The decay curves were fitted with a biexponential function, and the calculated average lifetimes are displayed.

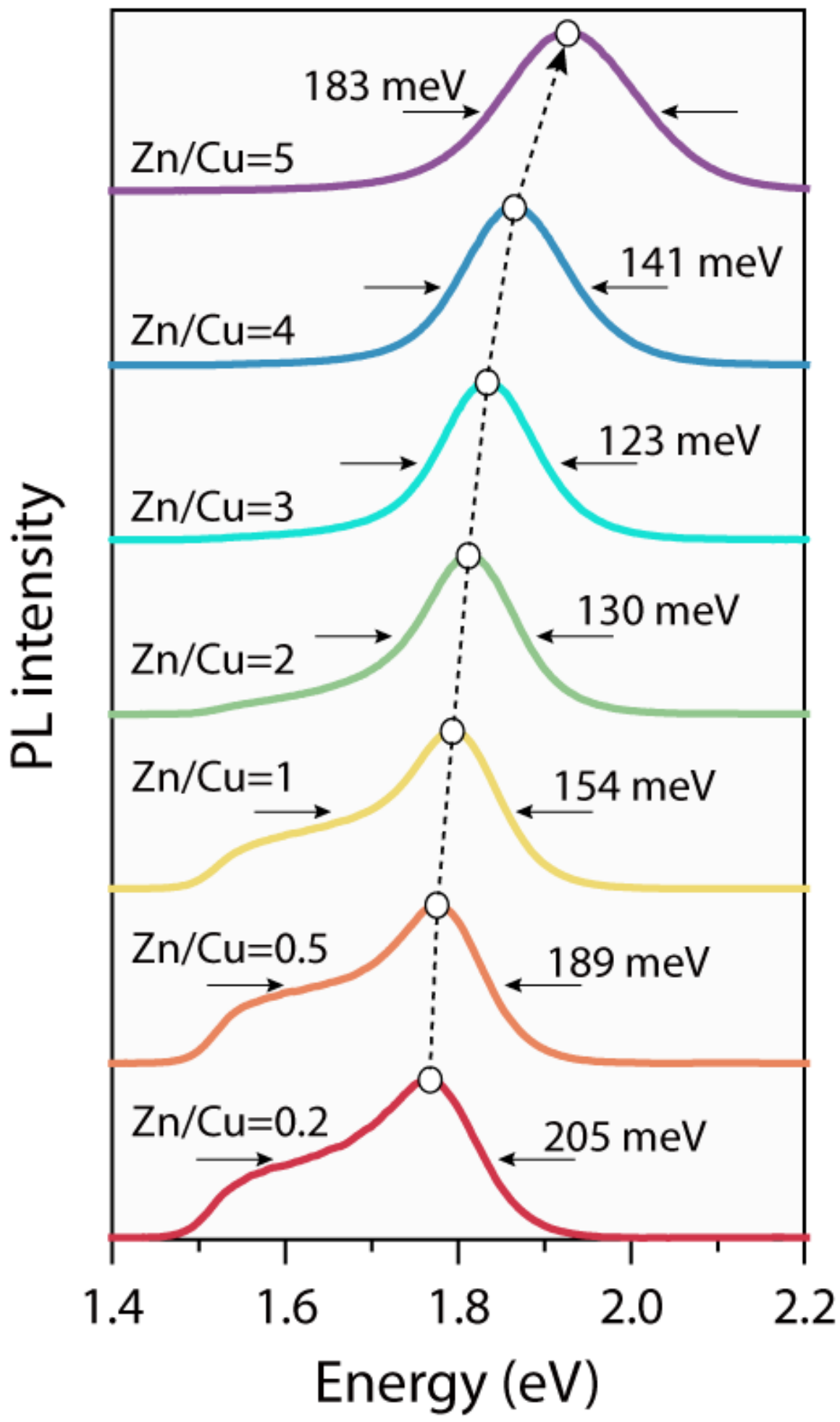

**Figure S6.** PL spectra of CIZS QDs with varying Zn/Cu feeding ratios.

**Structure, Strain, Crystallinity, and Composition**

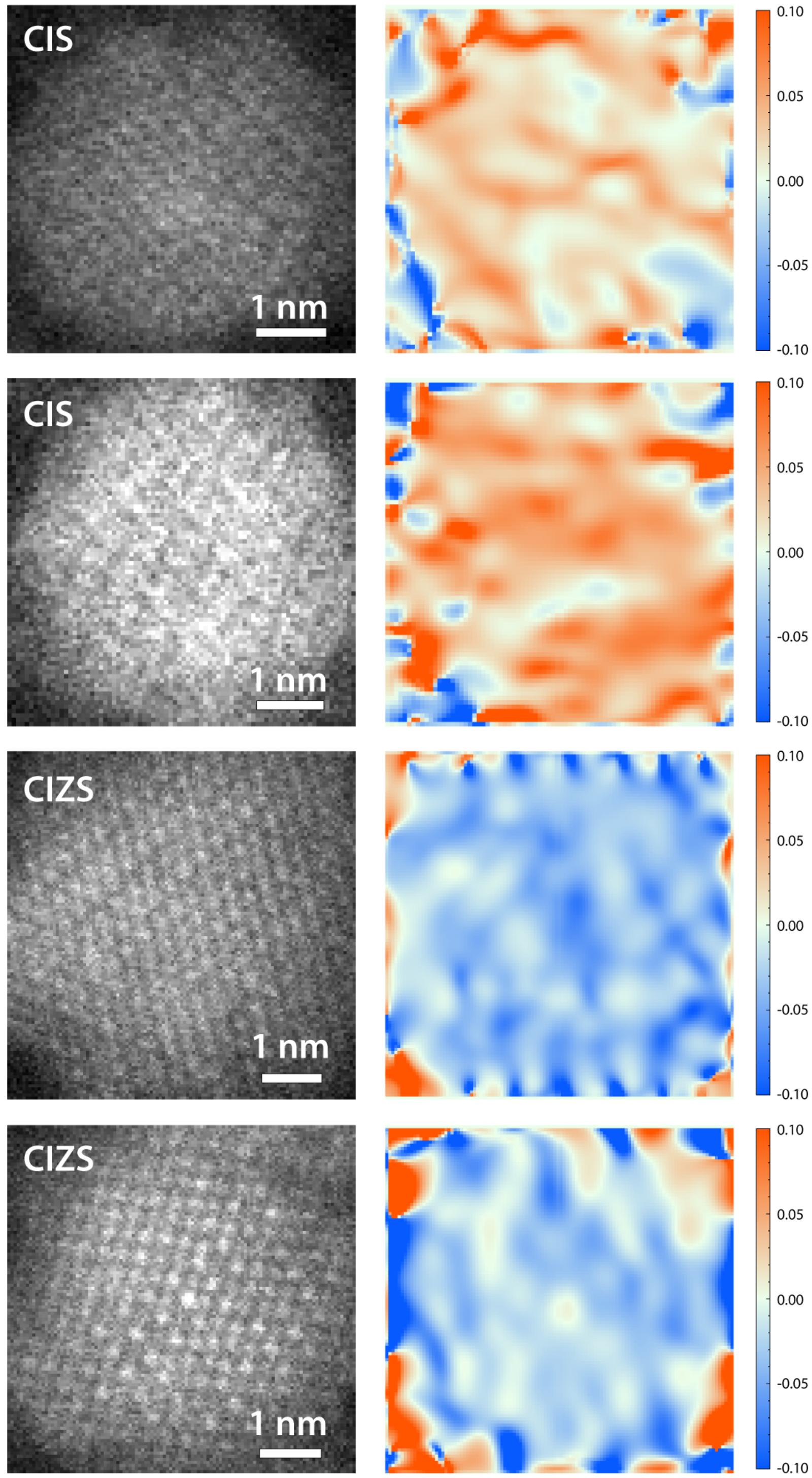

**Figure S7.** High-resolution HAADF-STEM images and corresponding strain distribution maps of CIS and CIZS QDs.

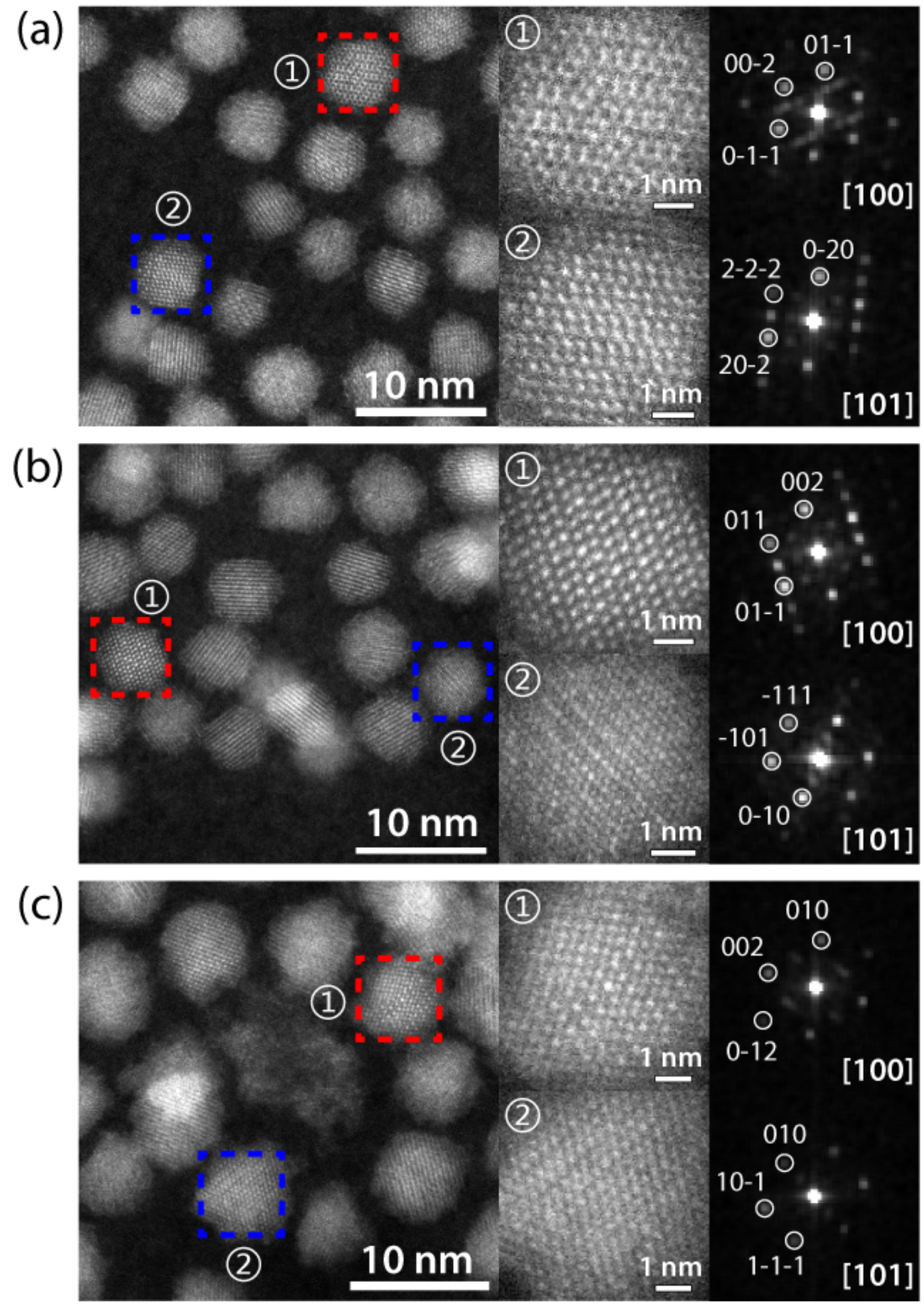

**Figure S8.** High-resolution HAADF-STEM images of CIZS **(a)**, CIZGS **(b)**, and CIZGS/ZnS QDs **(c)** viewed along the [100] and [101] zone axes, with corresponding FFT patterns.

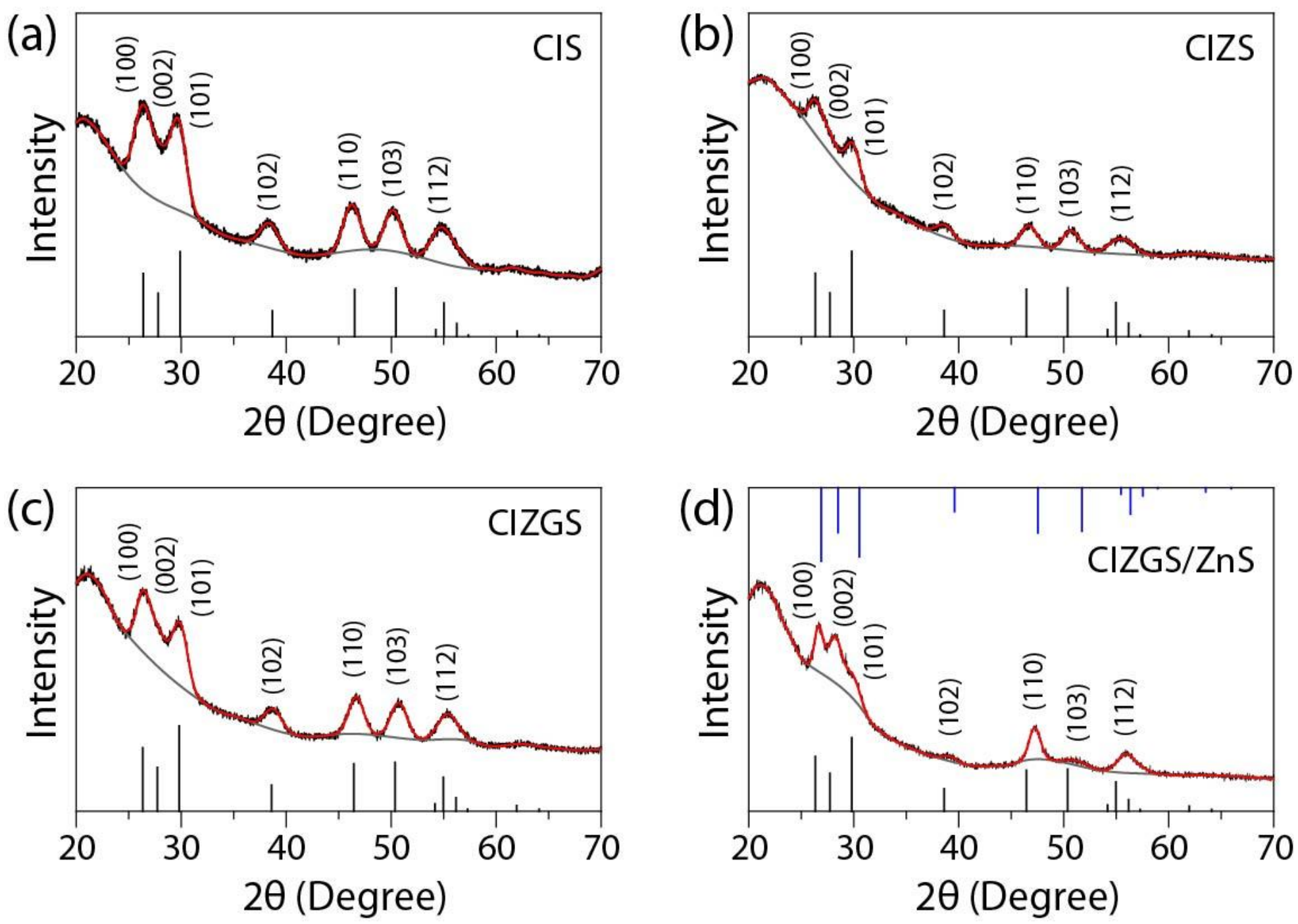

**Figure S9.** X-ray diffraction (XRD) patterns of CIS **(a)**, CIZS **(b)**, CIZGS **(c)**, and CIZGS/ZnS QDs **(d)**. The bottom black vertical lines in (a-d) and the top blue vertical lines in (d) correspond to the reference diffraction patterns for hexagonal wurtzite $CuInS_2$ (JCPDS card No. 97-016-3489) and hexagonal wurtzite ZnS (JCPDS card No. 04-008-7254), respectively. Background removal from the XRD patterns was performed using asymmetric least squares smoothing (grey curve), followed by multi-peak Gaussian fitting (red curve) for peak deconvolution. The integrated intensity and full width at half maximum (FWHM) values obtained from the Gaussian fitting were used to calculate the average crystallite size and crystallinity[10-12], with the results summarized in Table S2.

**Table S2.** The crystallite sizes (***D***) of CIS, CIZS, CIZGS, and CIZGS/ZnS QDs were estimated using Scherrer's equation ($D = K\lambda/(\beta\cos\theta)$[12], where $\boldsymbol{K}$ is the Scherrer constant (0.9), $\boldsymbol{\lambda}$ is the X-ray wavelength (0.15406 nm), $\boldsymbol{\beta}$ is the full width at half maximum (FWHM) in radians, and $\boldsymbol{\theta}$ is the diffraction peak position in radians. The values of $\boldsymbol{\theta}$ and $\boldsymbol{\beta}$ were determined by multi-peak Gaussian fitting as shown in Figure S9. Crystallinity was calculated as the ratio of the integrated area of all crystalline diffraction peaks to the total integrated area of the XRD pattern[13,14].

| CIS (Wurtzite-WZ) | | | | | |
|---|---|---|---|---|---|
| Plane | 2θ (Degree) | $\beta$ (Degree) | $D$ (nm) | Average $D$ (nm) | Crystallinity (%) |
| (100) WZ | 26.30 | 1.66 | 4.92 | 5.0 ± 0.3 | 78.4 |
| (002) WZ | 27.98 | 1.75 | 4.67 | | |
| (101) WZ | 29.74 | 1.54 | 5.35 | | |
| (102) WZ | 38.38 | 1.82 | 4.62 | | |
| (110) WZ | 46.23 | 1.66 | 5.20 | | |
| (103) WZ | 50.18 | 1.73 | 5.06 | | |
| (112) WZ | 54.52 | 1.83 | 4.88 | | |
| CIZS (CIS-Wurtzite-WZ) | | | | | |
| Plane | 2θ (Degree) | $\beta$ (Degree) | $D$ (nm) | Average $D$ (nm) | Crystallinity (%) |
| (100) WZ | 26.25 | 1.60 | 5.09 | 4.9 ± 0.5 | 81.7 |
| (002) WZ | 27.48 | 1.81 | 4.52 | | |
| (101) WZ | 29.91 | 1.60 | 5.13 | | |
| (102) WZ | 38.64 | 1.63 | 5.17 | | |
| (110) WZ | 46.70 | 1.62 | 5.35 | | |
| (103) WZ | 50.73 | 1.70 | 5.16 | | |
| (112) WZ | 55.50 | 2.19 | 4.09 | | |
| CIZGS (CIS-Wurtzite-WZ) | | | | | |
| Plane | 2θ (Degree) | $\beta$ (Degree) | $D$ (nm) | Average $D$ (nm) | Crystallinity (%) |
| (100) WZ | 26.31 | 1.55 | 5.25 | 4.9 ± 0.5 | 82.1 |
| (002) WZ | 27.63 | 1.87 | 4.38 | | |
| (101) WZ | 29.90 | 1.62 | 5.08 | | |
| (102) WZ | 38.76 | 1.70 | 4.95 | | |
| (110) WZ | 46.66 | 1.61 | 5.39 | | |
| (103) WZ | 50.75 | 1.71 | 5.14 | | |
| (112) WZ | 55.39 | 2.14 | 4.20 | | |
| CIZGS/ZnS (CIS-Wurtzite-WZ) | | | | | |
| Plane | 2θ (Degree) | $\beta$ (Degree) | $D$ (nm) | Average $D$ (nm) | Crystallinity (%) |
| (100) WZ | 26.65 | 1.23 | 6.66 | 5.7 ± 1.1 | 88.6 |
| (002) WZ | 28.23 | 1.51 | 5.42 | | |
| (101) WZ | 29.98 | 1.31 | 6.28 | | |
| (102) WZ | 39.10 | 1.51 | 5.57 | | |
| (110) WZ | 47.24 | 1.17 | 7.42 | | |
| (103) WZ | 51.24 | 2.13 | 4.13 | | |
| (112) WZ | 55.97 | 1.94 | 4.64 | | |

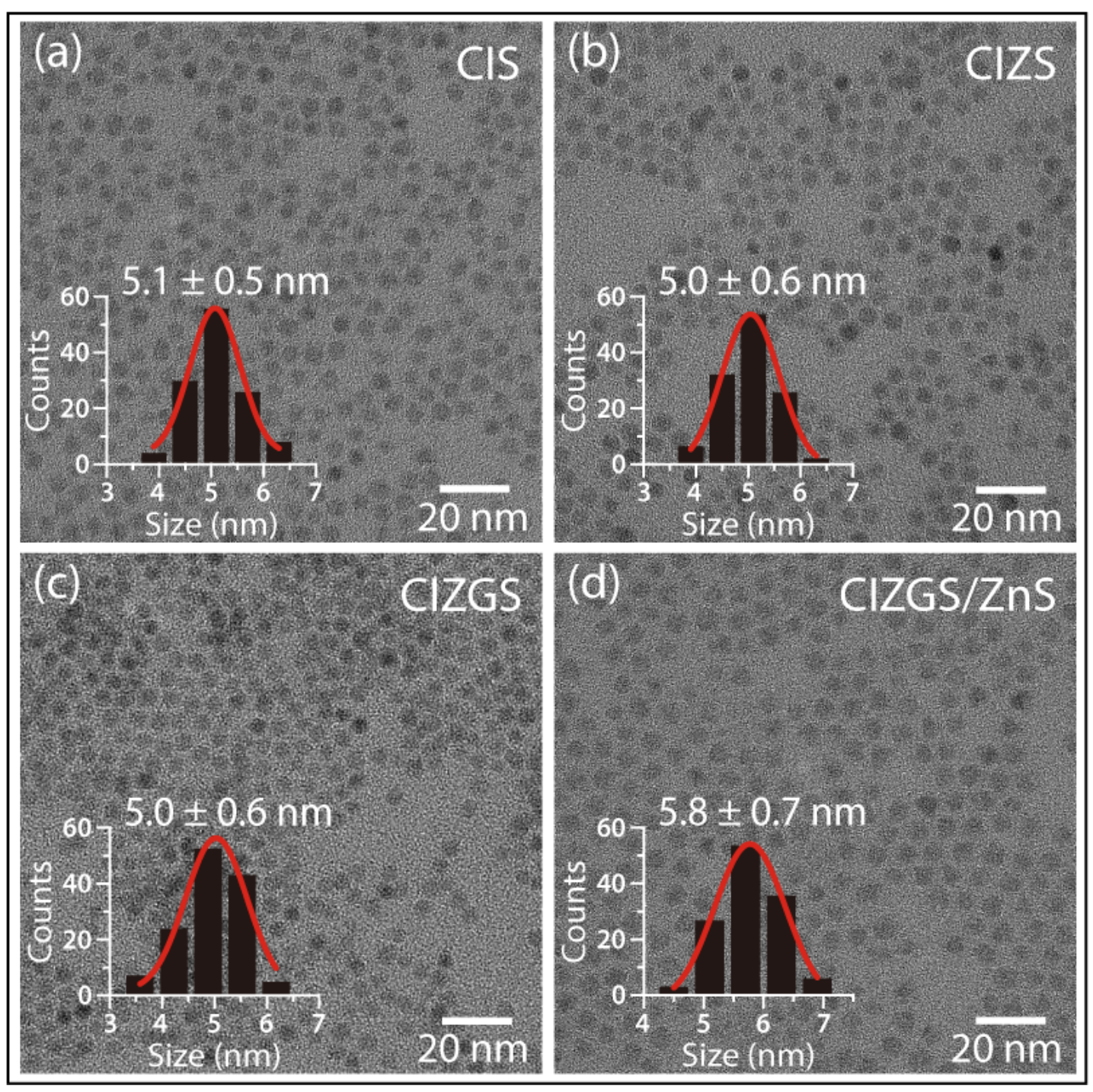

**Figure S10.** TEM images and corresponding size distribution histograms of CIS nanocrystals **(a)**, CIZS **(b)**, CIZGS **(c)**, and CIZGS/ZnS core/shell QDs **(d)**. The histograms were obtained by measuring over 100 particles from each image and fitting the data with a Gaussian distribution function.

**Table S3.** The composition of CIS, CIZS, CIZGS, and CIZGS/ZnS QDs was determined via three distinct techniques. Note: The ratios are normalized to Cu. For core/shell QDs where Cu is predominantly confined to the core and can be partially depleted/diluted during shelling, Cu-normalization can magnify the apparent Zn enrichment. Therefore, we use these ratios primarily to discuss qualitative trends and surface–bulk partitioning (XPS vs EDS), rather than absolute stoichiometry or shell thickness.

| Samples | ICP | | | XPS | | | EDS | | |
|---|---|---|---|---|---|---|---|---|---|
| | In/Cu | Zn/Cu | Ga/Cu | In/Cu | Zn/Cu | Ga/Cu | In/Cu | Zn/Cu | Ga/Cu |
| CIS | 0.77 | 0 | 0 | 0.67 | 0 | 0 | 0.68 | 0 | 0 |
| CIZS | 2.36 | 4.94 | 0 | 1.55 | 5.17 | 0 | 0.57 | 0.85 | 0 |
| CIZGS | 2.22 | 3.85 | 4.52 | 0.68 | 1.42 | 2.61 | 0.61 | 0.80 | 0.16 |
| CIZGS/ZnS | 1.43 | 21.27 | 5.15 | 0.25 | 9.80 | 1.54 | 0.69 | 2.14 | 0.27 |

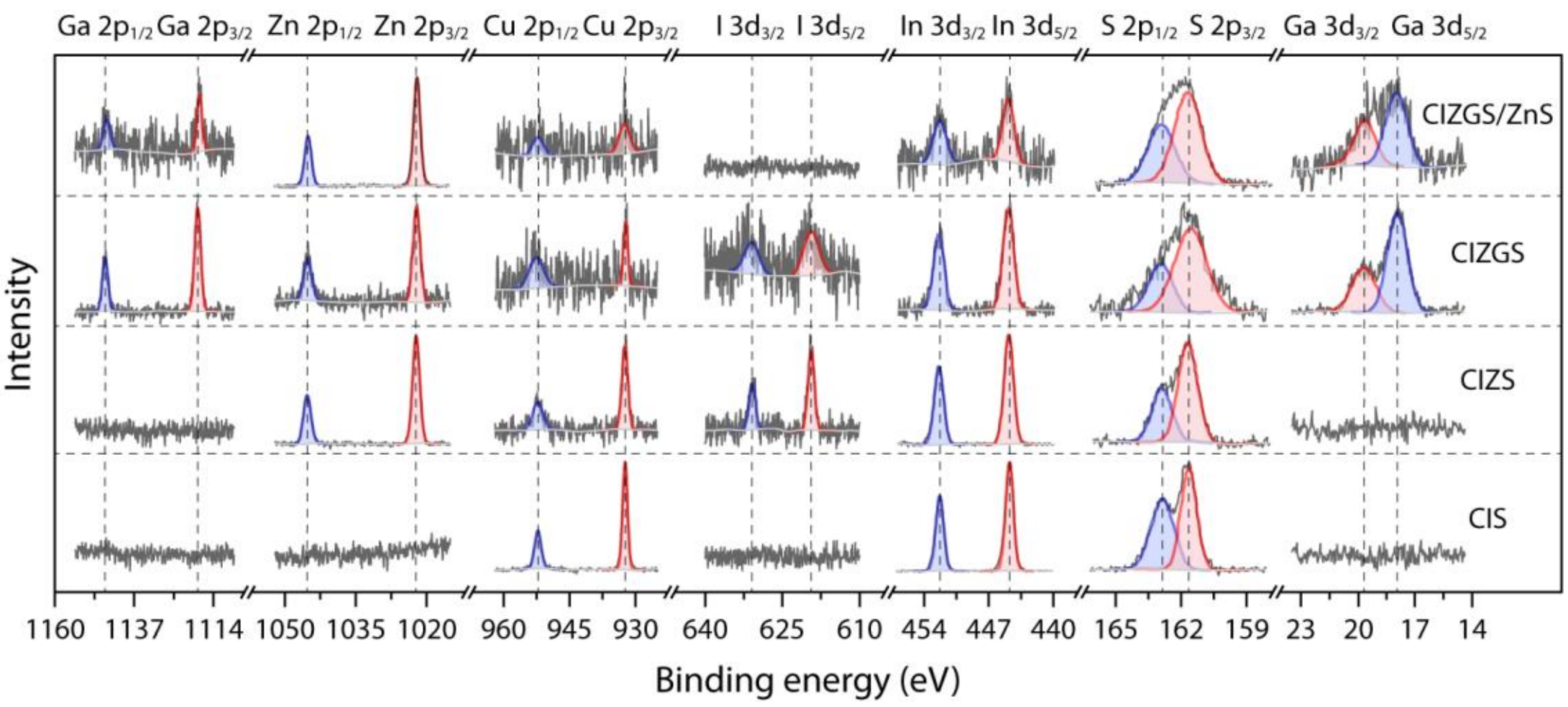

**Figure S11.** High-resolution XPS spectra of Ga, Zn, Cu, I, In, and S elements for CIS, CIZS, CIZGS, and CIZGS/ZnS QDs. Measured signals are shown in black, background fits in gray, and peak fittings in colored lines. The effective doping of $Ga^{3+}$ was confirmed by the distinct characteristic peaks in the Ga 2p and Ga 3d regions. The binding energy positions are consistent with the Ga(III) oxidation state, and no characteristic peaks attributable to a $Ga_2O_3$ phase were detected, indicating that Ga primarily exists in a doped state within the sulfide lattice.

## Kinetic Stabilization of the Defect Landscape by $Ga^{3+}$ Incorporation

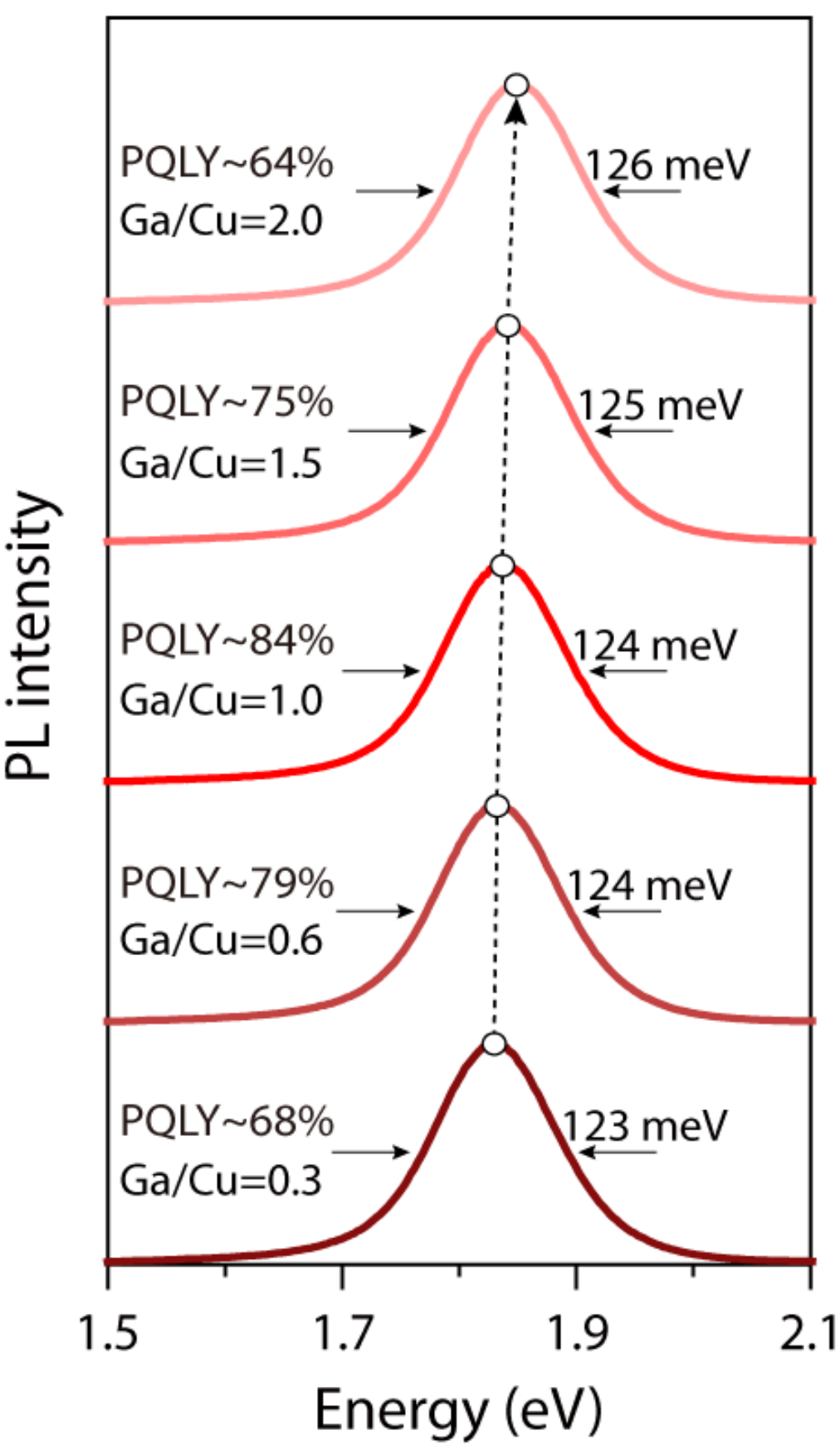

**Figure S12.** PL spectra of CIZGS QDs with varying Ga/Cu feeding ratios.

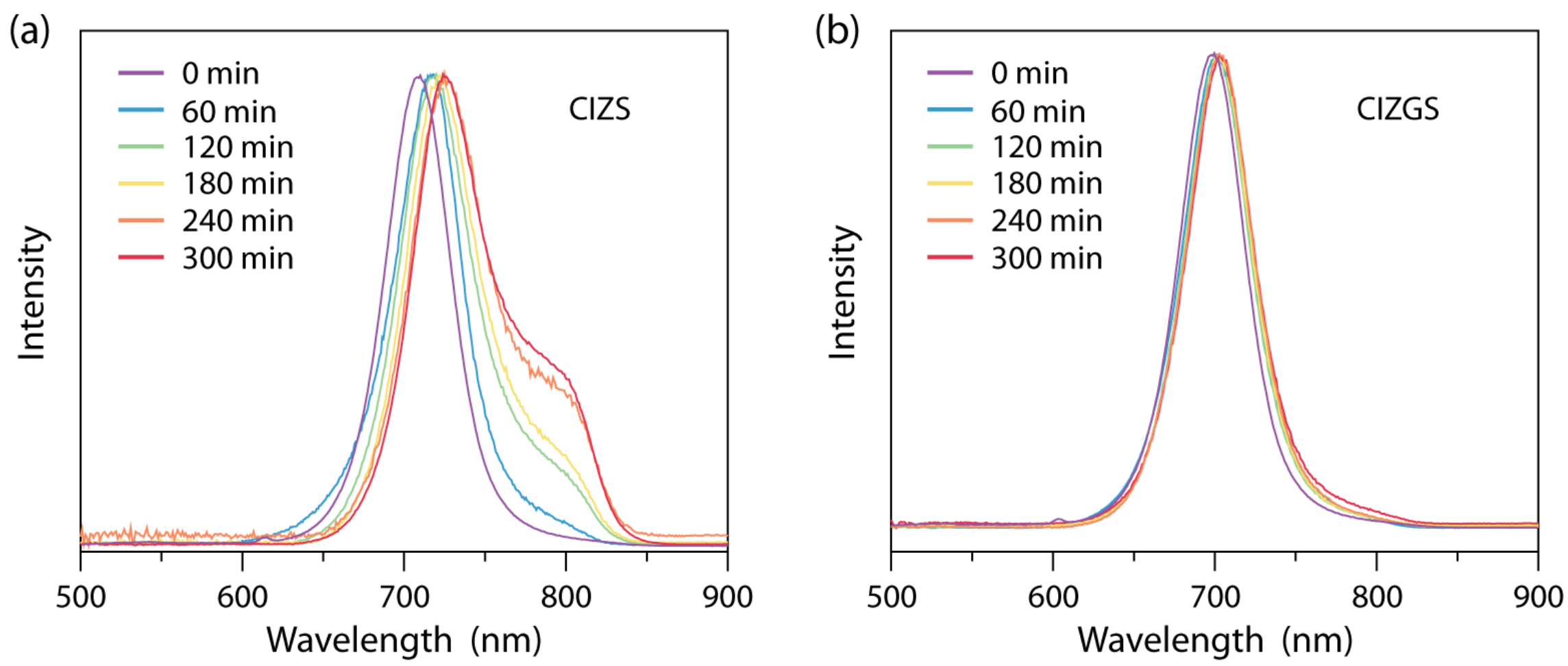

**Figure S13.** PL evolution of CIZS **(a)** and CIZGS **(b)** at 230 °C.

## Bandgap Analysis

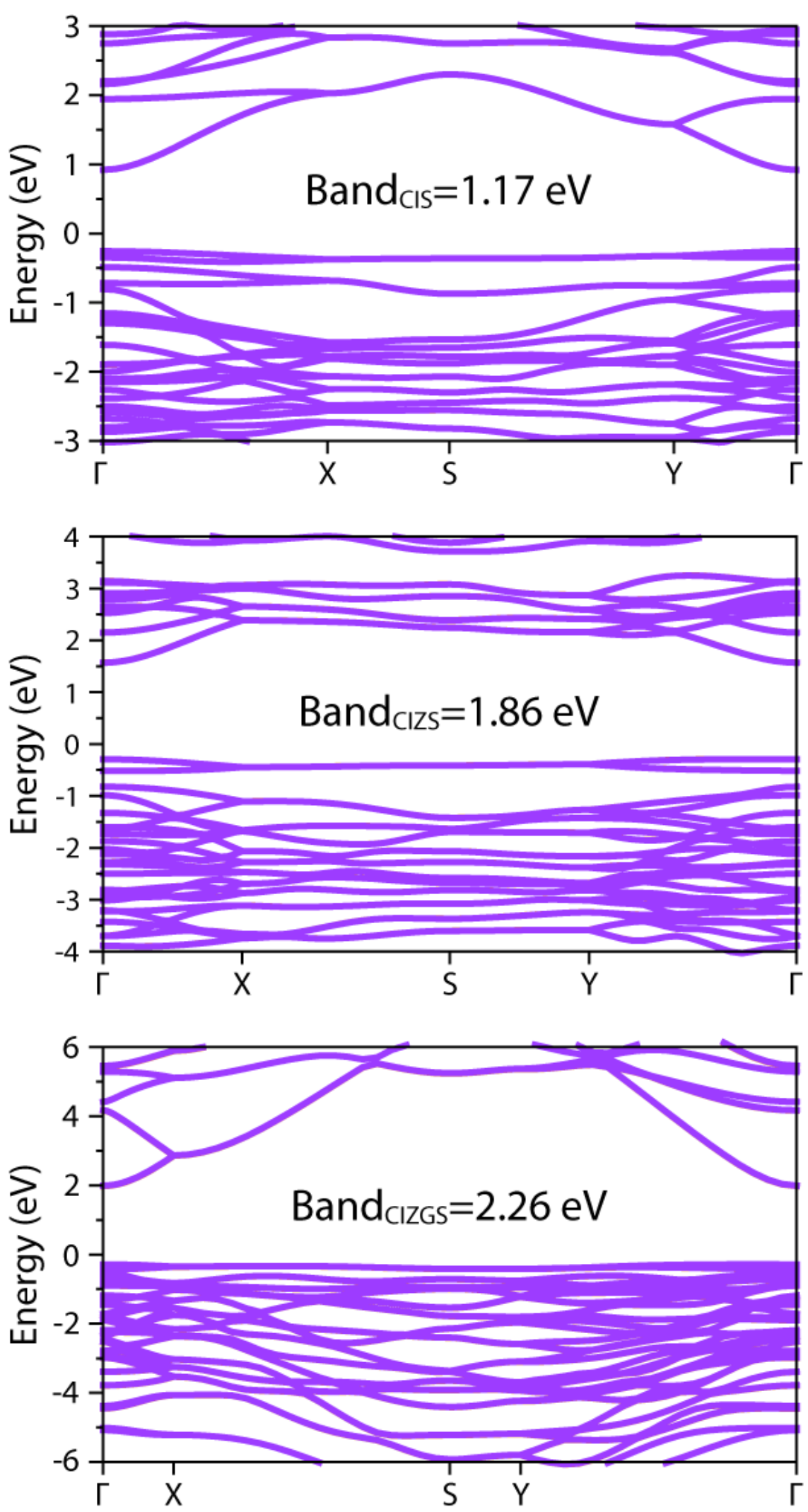

**Figure S14.** Band gap calculation of CIS, CIZS, and CIZGS.

## Single-Dot Spectroscopy and Photophysics Statistics

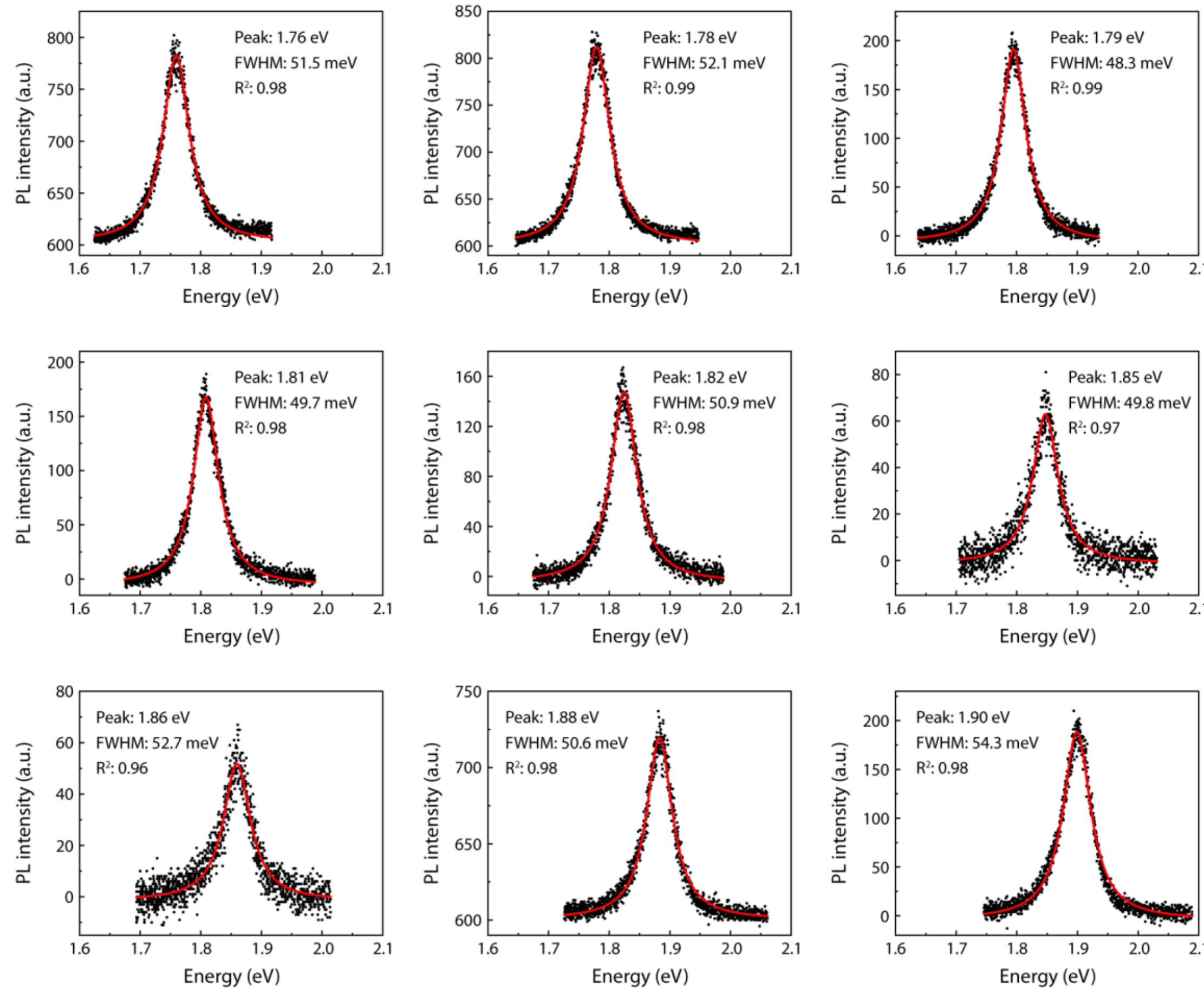

**Figure S15.** Randomly-selected single-dot spectra of CIZGS/ZnS QD at room temperature.

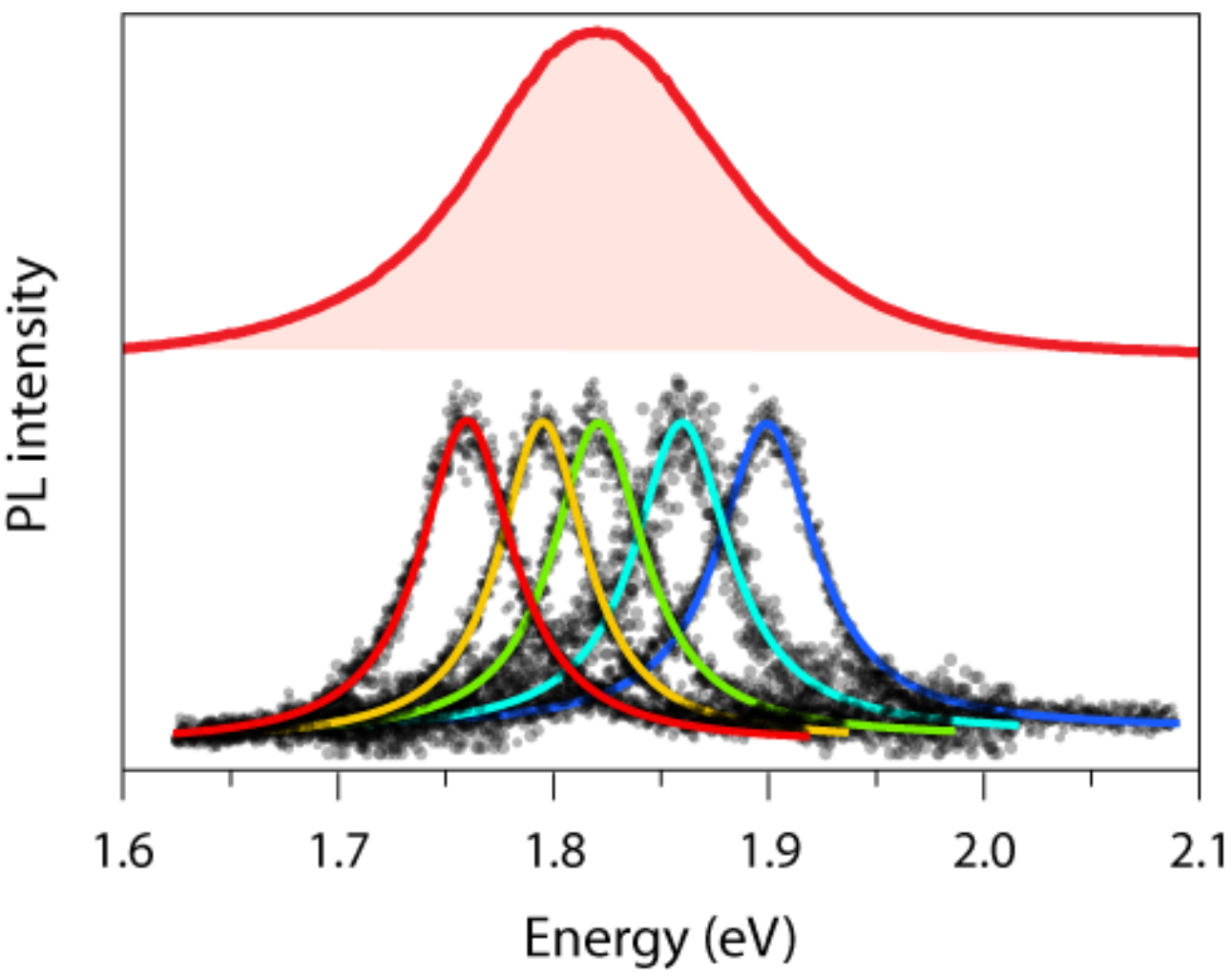

**Figure S16.** Comparison of the ensemble PL spectrum (red shaded area) with the scatter plot of PL spectra from individual QDs (gray circles), each fitted with a Lorentzian function.

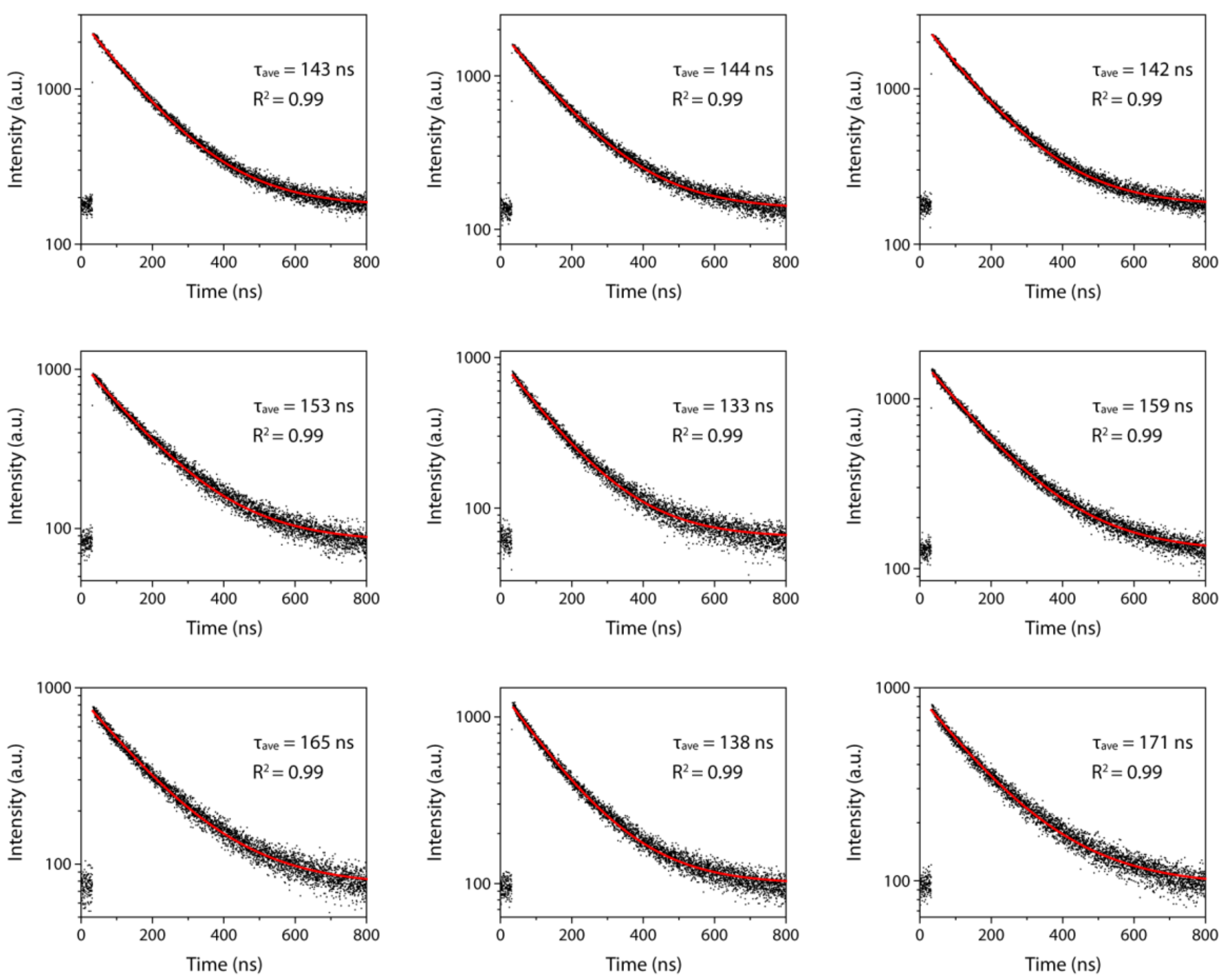

**Figure S17.** PL decay curve of a single CIZGS/ZnS QD, which was fitted with a mono-exponential function.

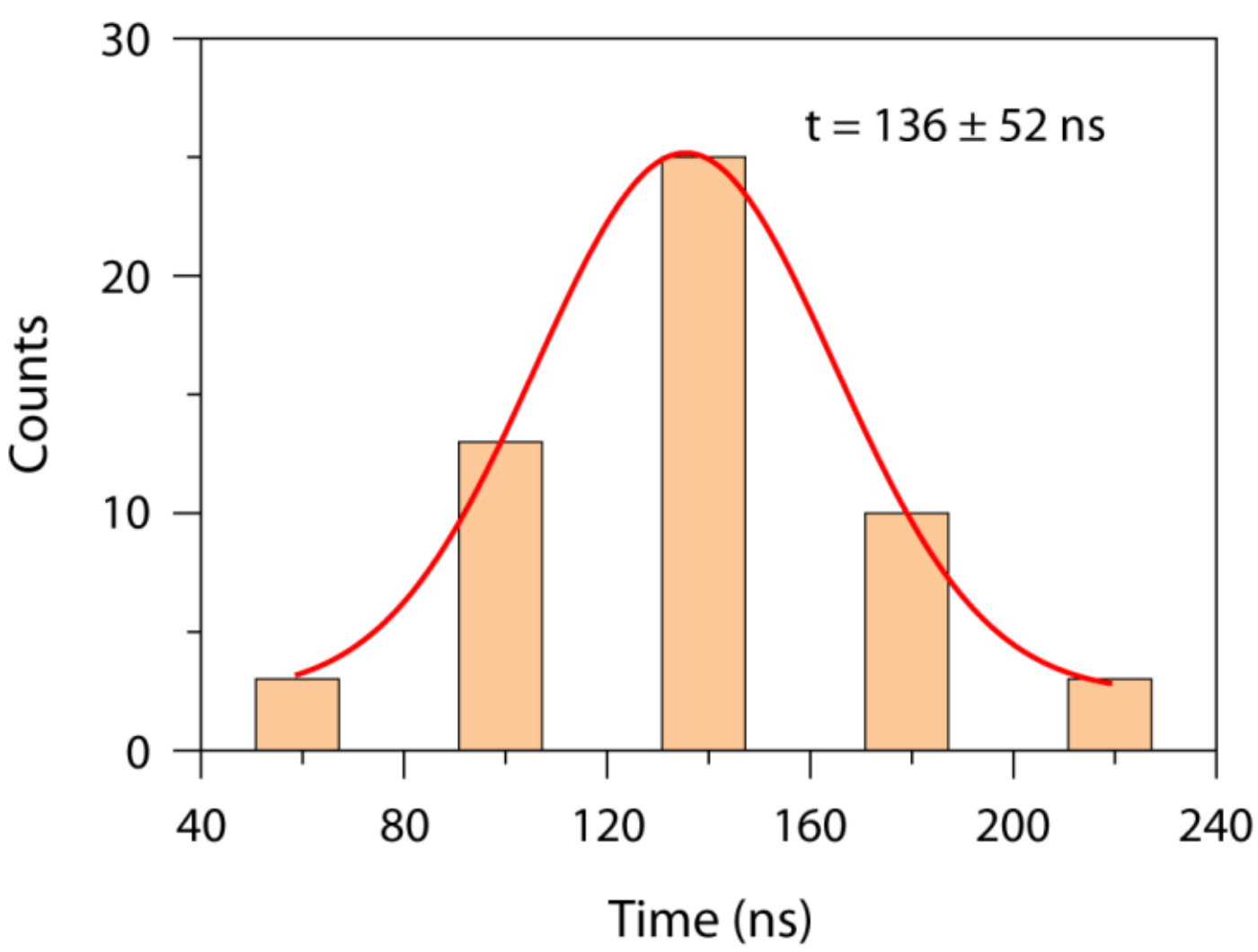

**Figure S18.** Statistical plot of the average lifetime obtained from mono-exponential fitting of the PL decay curves for 55 CIZGS/ZnS nanoparticles.

## LSC Device Characterization and Stability

**Table S4.** Performance summary of representative heavy-metal-free narrow-emitting QDs.

| QDs | PL peak (nm) | FWHM (nm) | FWHM (meV) | PLQY (%) | Ref |
|---|---|---|---|---|---|
| CIZGS/ZnS | 684 | ~50 | ~128 | 98 | this work |
| Cu–Ga–S | 475 | 29 | ~188 | 32 | [15] |
| Cu–Zn–In–Se | 1047 | ~139 | ~148 | 20 | [16] |
| CZISe/ZnS/$Al_2O_3$ | 1070 | ~137 | ~112 | 53.5 | [17] |
| Ag(In,Ga)$S_2$/Ag$GaS_2$ | ~517 | 30 | ~145 | 96 | [18] |
| Ag(In,Ga)$S_2$/Ga$S_y$ | 528 | 31 | 138 | > 90 | [19] |
| Carbon QDs | 507 | 30 | ~115 | 72 | [20] |
| InP/ZnS | 472 | 47 | ~264 | ~50 | [21] |
| InP/ZnSe/ZnS | 635 | ~38 | ~116 | ~74 | [22] |

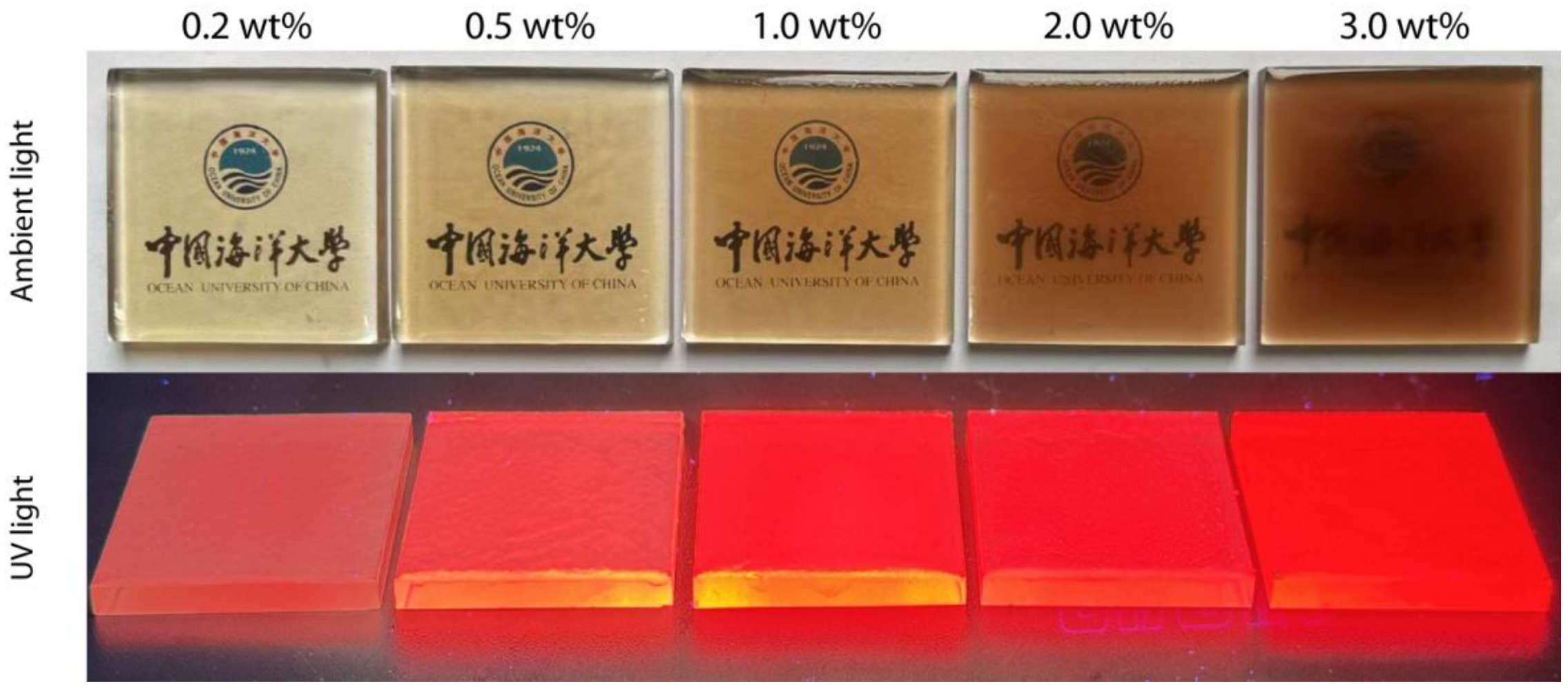

**Figure S19.** Photographs of LSCs doped with CIZGS/ZnS QDs at varying concentrations under ambient light and UV (405 nm) illumination. All devices have dimensions of 5×5×0.5 $cm^3$.

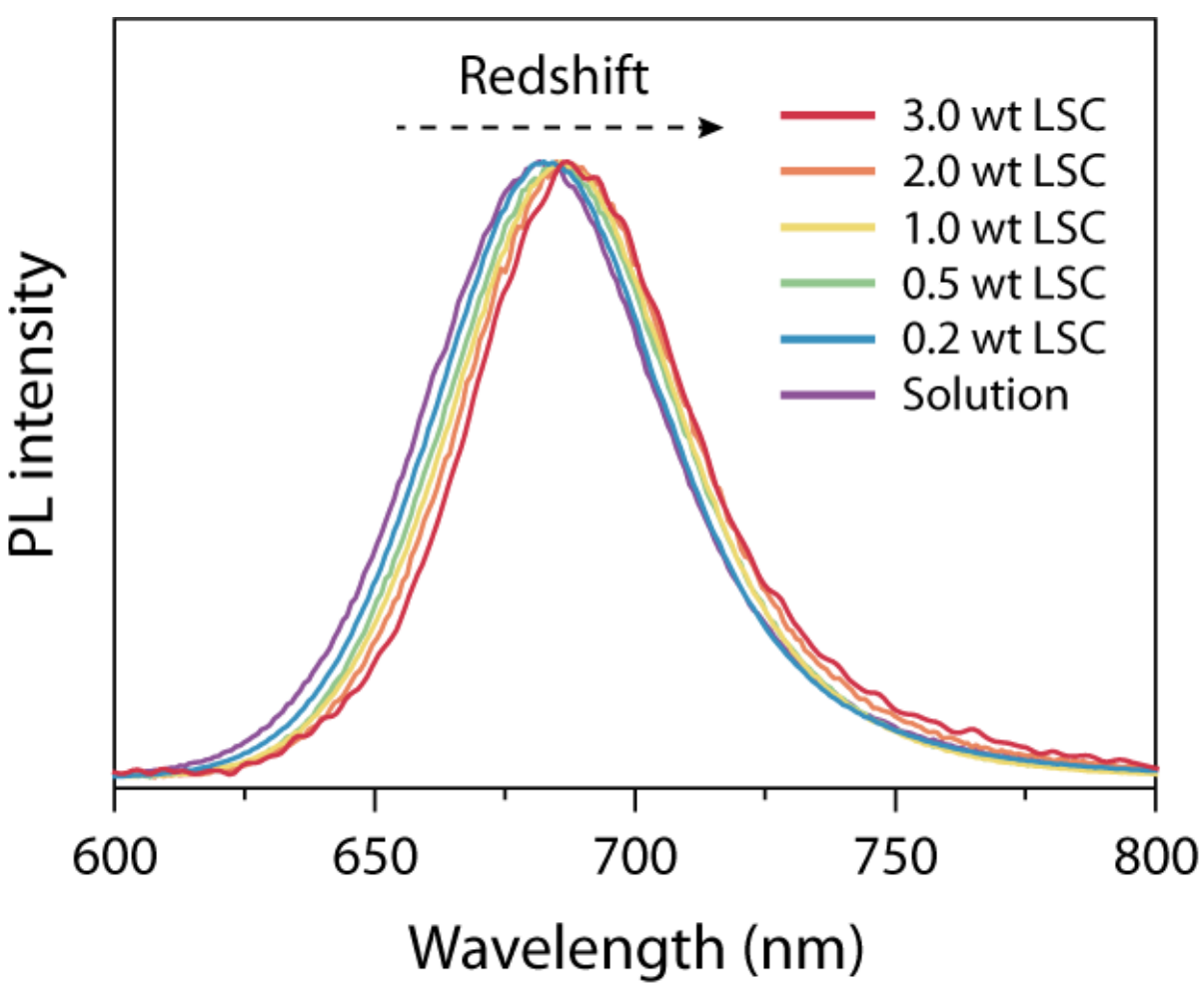

**Figure S20.** A comparison of the PL spectra of CIZGS/ZnS QDs in solution and within the LSC reveals a distinct red-shift in the LSC. This shift is accompanied by spectral broadening. These features are consistent with increased inter-dot coupling and/or partial aggregation during polymerization, which can enhance reabsorption and energy transfer in the solid matrix.

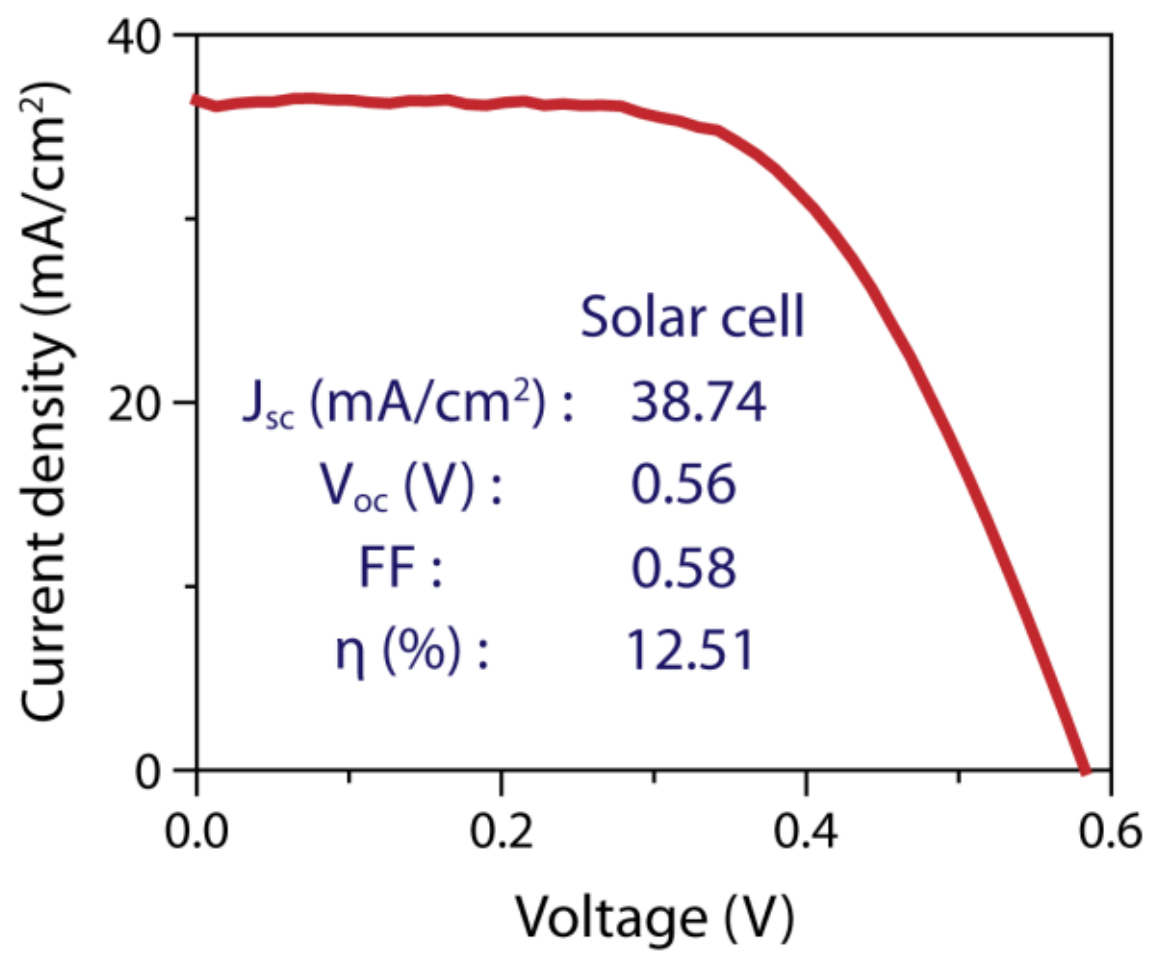

**Figure S21.** Characterization of the current-voltage characteristics of the silicon solar cells used for the luminescent solar concentrators.

**Table S5.** The characteristics, power conversion efficiencies (PCEs), and optical efficiencies ($\eta_{opt}$) of the as-fabricated LSCs in this work.

| Concentration (wt%) | Size ($cm^3$) | G | $P_0$ ($mW/cm^2$) | $J_{sc}$ ($mA/cm^2$) | $V_{oc}$ (V) | FF (%) | $\eta_{opt}$ (%) | PCE (%) |
|---|---|---|---|---|---|---|---|---|
| 0.0 | 5×5×0.5 | 2.5 | 100 | 0.70 | 0.41 | 0.58 | 0.72 | 0.17 |
| 0.2 | | | | 1.95 | 0.48 | 0.69 | 2.14 | 0.65 |
| 0.5 | | | | 3.77 | 0.51 | 0.69 | 4.14 | 1.33 |
| 1.0 | | | | 6.03 | 0.53 | 0.71 | 6.62 | 2.23 |
| 2.0 | | | | 10.04 | 0.53 | 0.68 | 11.02 | 3.62 |
| 3.0 | | | | 11.55 | 0.54 | 0.69 | 12.68 | 4.29 |

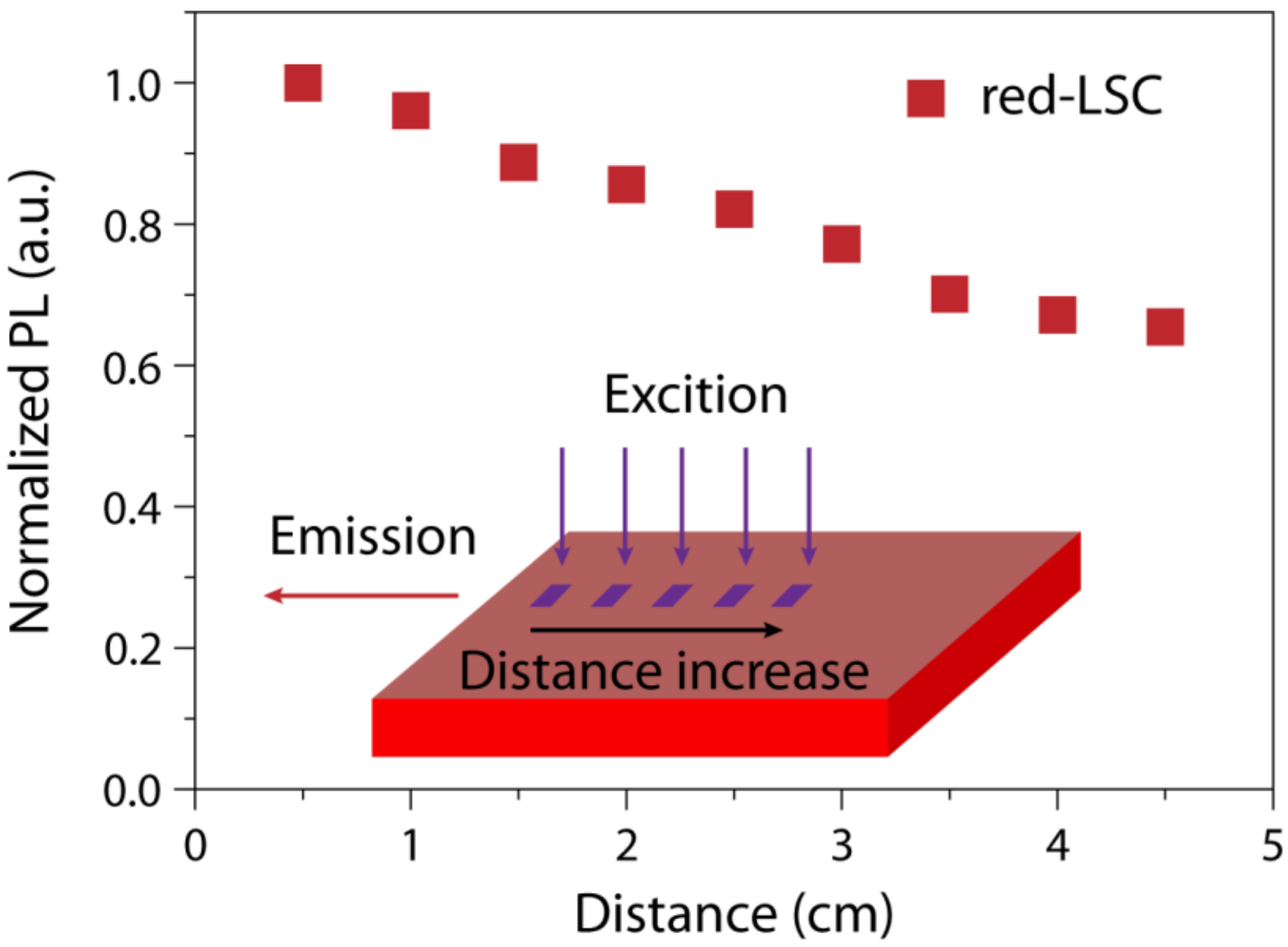

**Figure S22.** Spectral integrated intensity as a function of the distance between the excitation point and the edge of the LSC; the inset shows a schematic of the measurement setup.

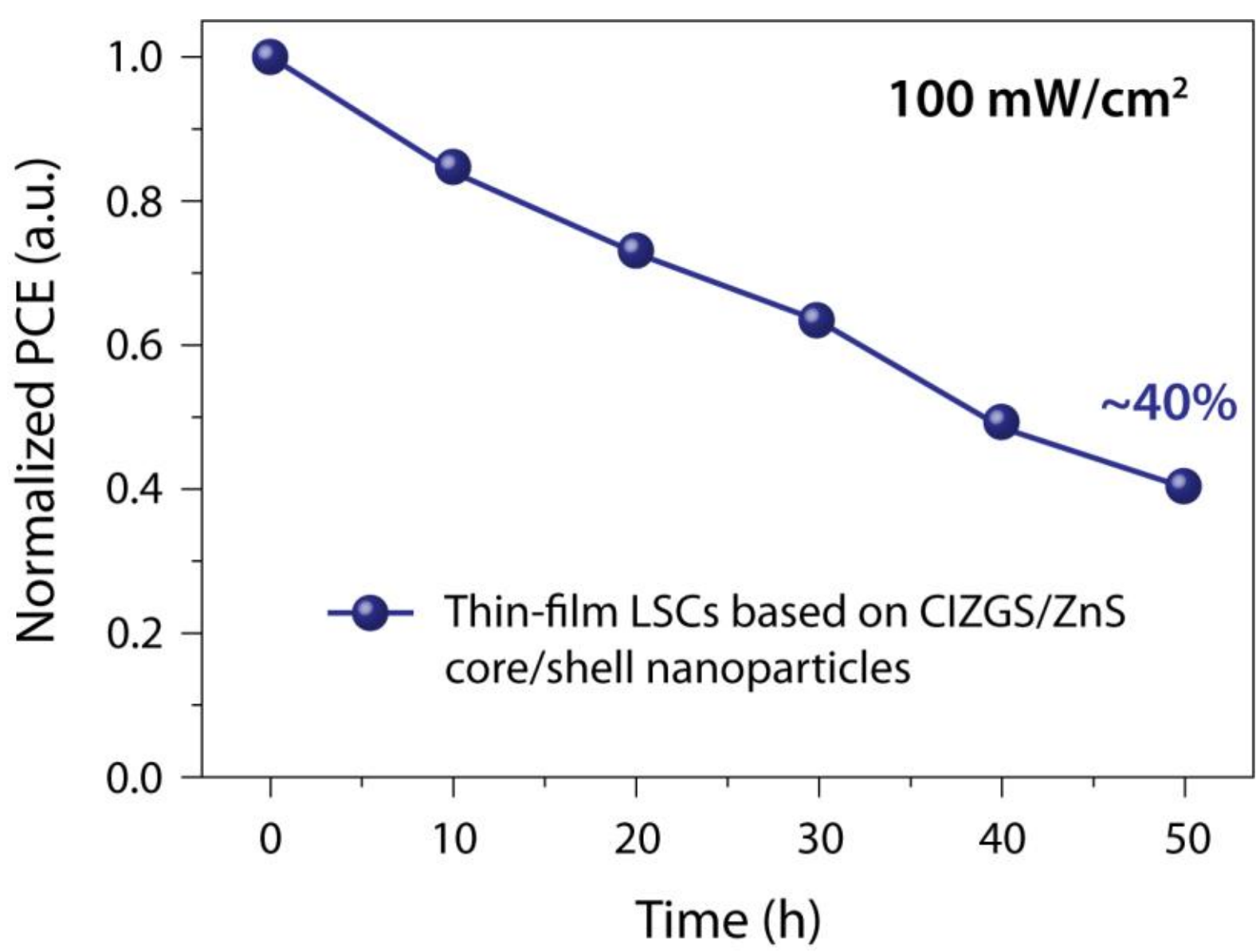

**Figure S23.** Operational stability of the CIZGS/ZnS QD‐based LSC under standard xenon lamp illumination (100 mW/cm$^2$).

**Table S6.** Performance comparison of various QD-based luminescent solar concentrators.

| Fluorophores | Waveguide | Size ($cm^3$) | PCE (%) | $\eta_{opt}$ (%) | Ref |
|---|---|---|---|---|---|
| CIZGS/ZnS | Glass | 5×5×0.5 | 0.65—4.29 | 2.14—12.68 | this work [b] |
| CIGS/ZnS | PMMA | 5×5×0.5 | 3.87 | \ | 23 [b] |
| Zn and Al co-doped CIS | Glass | 1.8×1.8×0.11 | 3.18 | 6.97 | 24 [b] |
| ZnCuInSe/ZnSe | PMMA | 10×10×0.5 | 0.44 | 3.38 | 8 [a] |
| CuGaAlS/ZnS | PLMA | 10×10×0.15 | 4.29 | \ | 25 [b] |
| $AgInS_2$/ZnS | PLMA | 10.4×10.4×0.2 | \ | \ | 26 [b] |
| $CsPbBr_3$@$SiO_2$ | PDMS | 2.5×2.5×0.2 | \ | 3.70—10.92 | 27 [b] |
| Silicon-Carbon | Glass | 5×5×0.2 | 4.36 | \ | 28 [b] |
| Carbon QDs | PMMA | 10×10×0.52 | 2.41 | 4.81 | 29 [a] |
| Graphene QDs | Glass | 10×10×0.63 | 1.4 | \ | 30 [b] |

[a]Measured under natural sunlight. [b]Measured under AM 1.5G (100 $mW/cm^2$) sunlight.